\documentclass[journal=nalefd,manuscript=article]{achemso}
\usepackage{color}

\usepackage[version=3]{mhchem} 

\author{Jia-Bin You}
\affiliation{%
Institute of High Performance Computing, Agency for Science, Technology, and Research (A*STAR), 1 Fusionopolis Way, \#16-16 Connexis, Singapore 138632.
}
\altaffiliation{These two authors contributed equally.}

\author{Xiao Xiong}
\affiliation{%
Institute of High Performance Computing, Agency for Science, Technology, and Research (A*STAR), 1 Fusionopolis Way, \#16-16 Connexis, Singapore 138632.
}
\altaffiliation{These two authors contributed equally.}

\author{Ping Bai}
\affiliation{%
Institute of High Performance Computing, Agency for Science, Technology, and Research (A*STAR), 1 Fusionopolis Way, \#16-16 Connexis, Singapore 138632.
}

\author{Zhang-Kai Zhou}
\affiliation{%
State Key Laboratory of Optoelectronic Materials and Technologies, School of Physics, Sun Yat-sen University, Guangzhou 510275, P. R. China.
}

\author{Ren-Min Ma}
\affiliation{%
State Key Laboratory for Mesoscopic Physics and Collaborative Innovation Center of Quantum Matter, School of Physics, Peking University, Beijing 100871, P. R. China.
}

\author{Wan-Li Yang}
\affiliation{%
State Key Laboratory of Magnetic Resonance and Atomic and Molecular Physics, Wuhan Institute of Physics and Mathematics, Chinese Academy of Sciences, Wuhan 430071, P. R. China.
}

\author{Yu-Kun Lu}
\affiliation{%
State Key Laboratory for Mesoscopic Physics and Collaborative Innovation Center of Quantum Matter, School of Physics, Peking University, Beijing 100871, P. R. China.
}

\author{Yun-Feng Xiao}
\affiliation{%
State Key Laboratory for Mesoscopic Physics and Collaborative Innovation Center of Quantum Matter, School of Physics, Peking University, Beijing 100871, P. R. China.
}

\author{Ching Eng Png}
\affiliation{%
Institute of High Performance Computing, Agency for Science, Technology, and Research (A*STAR), 1 Fusionopolis Way, \#16-16 Connexis, Singapore 138632.
}

\author{Francisco J. Garcia-Vidal}
\email{fj.garcia@uam.es}
\affiliation{%
Departamento de Fisica Teorica de la Materia Condensada and Condensed Matter Physics Center (IFIMAC), Universidad Autonoma de Madrid, E-28049 Madrid, Spain.
}

\author{Cheng-Wei Qiu}
\email{chengwei.qiu@nus.edu.sg}
\affiliation{%
Department of Electrical and Computer Engineering, National University of Singapore, 4 Engineering Drive 3, Singapore 117583.
}

\author{Lin Wu}
\email{wul@ihpc.a-star.edu.sg}
\affiliation{%
Institute of High Performance Computing, Agency for Science, Technology, and Research (A*STAR), 1 Fusionopolis Way, \#16-16 Connexis, Singapore 138632.
}

\title{Reconfigurable photon sources based on quantum plexcitonic systems
}

\begin{document}

\begin{abstract}
A single photon in a strongly nonlinear cavity is able to block the transmission of the second photon, thereby converting incident coherent light into anti-bunched light, which is known as photon blockade effect.
On the other hand, photon anti-pairing, where only the entry of two photons is blocked and the emission of bunches of three or more photons is allowed, is based on an unconventional photon blockade mechanism due to destructive interference of two distinct excitation pathways.
We propose quantum plexcitonic systems with moderate nonlinearity to generate both anti-bunched and anti-paired photons.
The proposed plexitonic systems benefit from subwavelength field localizations that make quantum emitters spatially distinguishable, thus enabling a reconfigurable photon source between anti-bunched and anti-paired states via tailoring the energy bands.
For a realistic nanoprism plexitonic system, two schemes of reconfiguration are suggested: (i) the chemical means by partially changing the type of the emitters; or (ii) the optical approach by rotating the polarization angle of the incident light to tune the coupling rate of the emitters.
These results pave the way to realize reconfigurable nonclassical photon sources in a simple quantum plexcitonic platform with readily accessible experimental conditions.
\end{abstract}

\section{Introduction}
Generation and manipulation of nonclassical light lie at the heart of quantum science and technology. For example, quantum key distribution and quantum random number generation rely on single photons for secure quantum communication \cite{RevModPhys.81.1301,RevModPhys.89.015004}, while quantum metrology and sensing exploit wave packets of a fixed number of photons to achieve enhanced sensitivity and energy efficiency \cite{RevModPhys.90.035005,doi:10.1117/1.OE.53.8.081910}.
Conventionally, such single photon sources can be realized when a quantum emitter is strongly coupled to a cavity,
in which the optical nonlinearity is present at a single photon level.
The resulting energy levels form the anharmonic Jaynes-Cummings ladder, which gives rise to conventional photon blockade (PB). In PB, an emitter in a cavity effectively modifies the cavity resonance after a single photon is absorbed, preventing subsequent photons to pass through and creating an anti-bunched single photon stream.
This PB effect has been realized in quantum dot-cavity systems \cite{Dayan1062},
spin ensemble-cavity systems \cite{saez2017enhancing,saez2018photon}, Kerr-nonlinearity cavities \cite{PhysRevLett.79.1467}, transmission line resonators \cite{PhysRevLett.106.243601}, and optomechanical systems
\cite{PhysRevLett.107.063601}.
The signature of PB is usually observed by measuring the second-order correlation function at zero delay $g^{(2)}(0)$. The value of $g^{(2)}(0)<1$, manifesting a sub-Poissonian photon statistics, indicates that the system is in PB regime.

Recently, an alternative route is found to achieve photon blockade for generating nonclassical light in cavity quantum electrodynamic systems \cite{PhysRevLett.121.043601,PhysRevLett.121.043602,PhysRevLett.121.133601,PhysRevLett.121.153601,Jing2019}.
It is  predicted that the ``strong coupling'' condition to attain PB could be relaxed if more than one cavity or emitter is employed \cite{PhysRevLett.104.183601,PhysRevA.96.011801}.
This new mechanism, known as unconventional photon blockade (UPB), relies on the destructive quantum interference between different excitation pathways \cite{PhysRevA.83.021802,Carreno2016}.
The UPB should be jointly characterized by the second- and third-order correlation functions at zero delay, which requires $g^{(2)}(0)<1$ and $g^{(3)}(0)>1$.
Unlike the conventional PB, UPB suppresses the emission of two photons only, and allows the emission of single photon, as well as bunches of three or more photons simultaneously.
The realization of UPB has been extensively proposed in optomechanical systems \cite{0953-4075-46-3-035502},
quantum dot-cavity systems \cite{PhysRevLett.108.183601}, Kerr-nonlinearity cavities \cite{PhysRevA.92.023838}, and weakly nonlinear photonic molecules \cite{PhysRevA.90.033809}.
In light of these efforts, UPB could not only advance the development of single photon sources, but also enable the multiphoton emission that probably unveils richer physics \cite{PhysRevA.96.011801}.

\begin{figure}[!ht]
\centering
\includegraphics[scale=0.5]{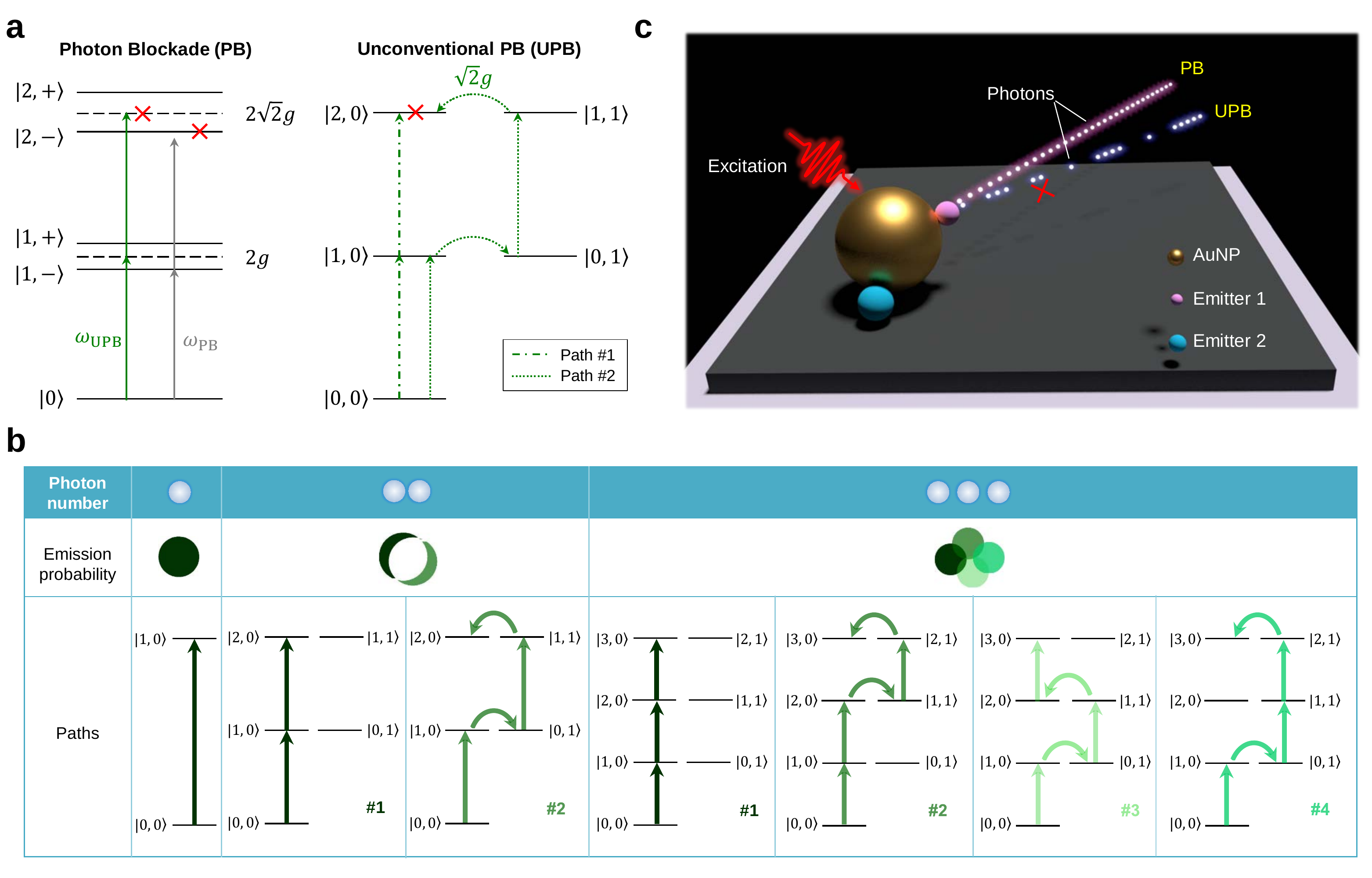}
\caption{
\textbf{Quantum plexcitonic system for a reconfigurable single photon source.}
(a) Energy level diagram showing a two-level quantum emitter coupled  to a cavity field in resonance, with coupling rate $g$. The suppression of two-photon emission is due to the anharmonic Jaynes-Cummings ladder for photon blockade (PB), and the destructive interference of two distinct excitation pathways for unconventional photon blockade (UPB).
(b) Detailed excitation pathway analysis of UPB for one-, two-, and three-photon emission.
(c) Schematics of the proposed quantum plexcitonic system for a reconfigurable single photon source between  anti-bunched (PB) and anti-paired (UPB) states.}
\label{fig1}
\end{figure}

In this work, we explore the possibility to realize both photon blockade effects in the same Jaynes-Cummings type of quantum plexcitonic system; and more interestingly, to switch between PB and UPB freely.
Plexcitons refer to polaritonic modes that result from coherent coupling between plasmons ($i.e.$, collective electron oscillations in a nanoscale plasmonic cavity) and excitons ($i.e.$, excitation quanta of the emitter).
Recently, three individual experiments on different plexcitonic systems, $i.e.$, silver-bowtie/semiconductor-quantum-dots \cite{Santhosh2016}, gold-nanoparticle-on-mirror/dye-molecules \cite{Chikkaraddy2016},
and cuboid-Au@Ag-nanorod/J-aggregates \cite{PhysRevLett.118.237401}, have concurrently confirmed that room-temperature strong coupling between a single quantum emitter and a single plasmonic cavity is indeed feasible with reported coupling rates of 120 meV \cite{Santhosh2016}, 90 meV \cite{Chikkaraddy2016} and 78 meV \cite{PhysRevLett.118.237401}, respectively.
These experiments at a single emitter level successfully push plasmonics into the quantum regime, where the anharmonic energy structure is now in action as illustrated in Fig. \ref{fig1}a, making PB highly feasible.
In fact, the second-order correlation function of the photon emission statistics has been recently measured at room temperature by coupling a single molecule with a plasmonic nanocavity, and is found to be pump wavelength dependent, varying from $g^{(2)}(0)=$ 0.4 to 1.45 \cite{ojambati2019quantum}.
Moreover, UPB is also readily achievable in such a plexcitonic system by carefully engineering the quantum interference between different excitation pathways for multiple photons. Taking the two-photon case as an example elaborated in Fig. \ref{fig1}b, two distinct excitation pathways ($i.e.$, path $\#$1: $|{0,0}\rangle\rightarrow|{1,0}\rangle\rightarrow|{2,0}\rangle$ and path $\#$2: $|{0,0}\rangle\rightarrow|{1,0}\rangle\rightarrow|{0,1}\rangle\rightarrow|{1,1}\rangle\rightarrow|{2,0}\rangle$) could destructively interfere and thus the two-photon emission is suppressed.
Here, $|i,j\rangle$ represents a plexcitonic state, where $i$ and $j$ correspond to the number of plasmons and excitons, respectively.

Compared to the photon blockade in photonic microcavity systems \cite{Hood1447,RevModPhys.73.565,Chiorescu2004,Reithmaier2004}, the plexcitonic nanostructures provide subwavelength extreme localization of plasmon fields beyond the diffraction limit \cite{doi:10.1021/acsphotonics.7b00674}, where the strong light-matter interaction is easier to be realized. More essentially, even the same type of emitters are involved, they can still be spatially distinguishable by the plasmon fields distributions, leading to greater degree of freedom to design a reconfigurable nonclassical photon source.
Due to the relatively large resonant and transition energies  for both plasmons and excitons ($i.e.$, $\hbar\omega\sim$ 2 eV), the mean thermal photon number \cite{PhysRevB.88.134511}, $n_{\text{th}}=[\exp{(\hbar\omega/k_{\text{B}}T)}-1]^{-1}$, is nearly zero at room temperature ($T\sim300$ K) with thermal energy $k_{\text{B}}T\sim$ 0.026 eV. Thus the plexcitonic system is more robust against temperature, in contrast to the cryogenic-temperature requirement for photonic microcavity systems.

Our proposed quantum plexcitonic system is schematically illustrated in Fig. \ref{fig1}c, which operates as a reconfigurable single photon source driven by coherent light.
The plexcitonic system consists of an Au nanoparticle (AuNP) plasmonic nanocavity (resonant energy $\omega_{\textrm{c}}$, decay rate $\kappa$) and distinguishable quantum emitters (transition energy $\omega_{\textrm{e1}}$ or $\omega_{\textrm{e2}}$, decay rate $\gamma_{\textrm{e1}}$ or $\gamma_{\textrm{e2}}$) with coupling rate of $g_\textrm{e1}$ or $g_{\textrm{e2}}$ to the plasmonic cavity.
By carefully designing this plexcitonic system, we are able to realize either PB or UPB effect.
As shown in Fig. \ref{fig1}c, PB generates anti-bunched single photon beam, whereas UPB generates anti-paired photons with certain probability to emit bunched photons of three or more.
It should be noted that a spherical AuNP is drawn in Fig. \ref{fig1}c for illustration purpose only.
In practice, plasmonic nanocavities offering strong field localization, such as bowtie \cite{Santhosh2016}, NP-on-mirror \cite{Chikkaraddy2016,xiong2019ultrastrong}, cuboid Au@Ag nanorod  \cite{PhysRevLett.118.237401} or nanoprism \cite{zengin2015realizing,cuadra2018observation} (shown later as our case study), would be more feasible.

\section{Results and discussions}
A full quantum mechanical description based on the Lindblad master equation is used to describe the quantum emitter, plasmonic cavity and the interaction between them (see Methods). Briefly, within the rotating frame \cite{Ste07,PhysRevA.100.053851}, our quantum plexcitonic system, driven by a weak coherent light (laser field $E_{l}=\kappa/50$ unless otherwise stated and laser frequency $\omega$), can be described by the following Hamiltonian:
\begin{equation}
\label{light_source}
\begin{split}
H&=\Delta_{\text{c}}a^{\dag}a+\Delta_{\textrm{e1}}\sigma_{\textrm{e1}}^{+}\sigma_{\textrm{e1}}^{-}+\Delta_{\textrm{e2}}\sigma_{\textrm{e2}}^{+}\sigma_{\textrm{e2}}^{-}+g_{\textrm{e1}}(a\sigma_{\textrm{e1}}^{+}+a^{\dag}\sigma_{\textrm{e1}}^{-})
+g_{\textrm{e2}}(a\sigma_{\textrm{e2}}^{+}+a^{\dag}\sigma_{\textrm{e2}}^{-})+E_{l}(a+a^{\dag}),\\
\end{split}
\end{equation}
where the first three terms represent the energies of the cavity and emitters,
with $\Delta_{\text{c}}=\omega_{\text{c}}-\omega$ and $\Delta_{\textrm{e}j}=\omega_{\textrm{e}j}-\omega$ ($j=1,2$) being the laser detunings for the cavity and emitters, respectively.
The cavity field is represented by bosonic operators with commutation relation $[a,a^{\dag}]=1$, and the emitters ($e.g.$, e$_1$ and e$_2$) are modeled as two-level excitonic systems $\omega_{\textrm{e}j}\sigma_{\textrm{e}j}^{+}\sigma_{\textrm{e}j}^{-}$.
The fourth and fifth terms represent the couplings between the cavity and the emitters via the Jaynes-Cummings interaction $g_{\textrm{e}j}(a\sigma_{\textrm{e}j}^{+}+a^{\dag}\sigma_{\textrm{e}j}^{-})$.
The last term corresponds to the coherent driving of the cavity.
Meanwhile, the plasmonic cavity and the emitters are exposed to dissipative environments, which are described by the density matrix $\rho$ following a Lindblad master equation:
\begin{equation}
\label{mseq_light_source}
\begin{split}
\partial_{t}{\rho}&=i[\rho,H]+\frac{\kappa}{2}\mathcal{D}[a]\rho+\frac{\gamma_{\textrm{e1}}}{2}\mathcal{D}[\sigma_{\textrm{e1}}^{-}]\rho+\frac{\gamma_{\textrm{e2}}}{2}\mathcal{D}[\sigma_{\textrm{e2}}^{-}]\rho,\\
\end{split}
\end{equation}
where $\kappa$ and $\gamma_{\textrm{e}j}$ are the decay rates of the cavity and the emitters, respectively. The Lindblad terms $\mathcal{D}[\hat{o}]\rho=2\hat{o}\rho\hat{o}^{\dag}-\rho\hat{o}^{\dag}\hat{o}-\hat{o}^{\dag}\hat{o}\rho$ with $\hat{o}=a$ or $\sigma_{\textrm{e}j}^{-}$, describe the different dissipation channels from either cavity or emitters to the environment.
We quantify the photon blockade effects by analyzing the equal-time second- and third-order correlations of the cavity field:
\begin{equation}
\label{second_correlation}
\begin{split}
g^{(2)}(0) = \frac{\langle a^{\dag2} a^{2}\rangle}{\langle a^{\dag} a\rangle^{2}},
\ \ \
g^{(3)}(0) = \frac{\langle a^{\dag3} a^{3}\rangle}{\langle a^{\dag} a\rangle^{3}},\\
\end{split}
\end{equation}
where the expectation value of an operator $\langle \hat{o} \rangle$ is obtained by $\langle \hat{o} \rangle = \textrm{Tr} [\rho\hat{o}$], with Tr denoting the trace of a matrix \cite{PhysRevLett.121.153601,PhysRevA.96.011801}.

In modern quantum optics, the equal-time correlation function $g^{(n)}(\tau=0)$ has been employed as a criterion to characterize the statistical and coherent properties of a light source, where $n$ is the order of the correlation function and $\tau$ is the time delay. For $g^{(n)}(0)>1$, $g^{(n)}(0)=1$, or $g^{(n)}(0)<1$, the $n$-photons are bunching, coherent, or anti-bunching, following super-Poissonian, Poissonian, or sub-Poissonian statistics, respectively \cite{PhysRevLett.121.153601,PhysRevA.96.011801}. By comparing any light source against a coherent light source with Poissonian distribution, it can be known that the $n$-photon emission is enhanced when $g^{(n)}(0)>1$ (classical light) and suppressed when $g^{(n)}(0)<1$ (quantum light). In this work, we use the criterion of suppressed $g^{(2)}(0)<1$ and $g^{(3)}(0)<1$ correlations to indicate the conventional photon blockade with sub-Poissonian photon distribution. On the other hand, suppressed $g^{(2)}(0)<1$ but enhanced $g^{(3)}(0)>1$ define the unconventional photon blockade\cite{PhysRevLett.121.153601,PhysRevA.96.011801,Carreno2016}, where two-photon emission is suppressed but the enhanced third-order correlation implies the emission of multiple photons.

\subsection*{Plexcitonic systems with indistinguishable emitters}

\begin{figure}[!ht]
\centering
\includegraphics[scale=0.45]{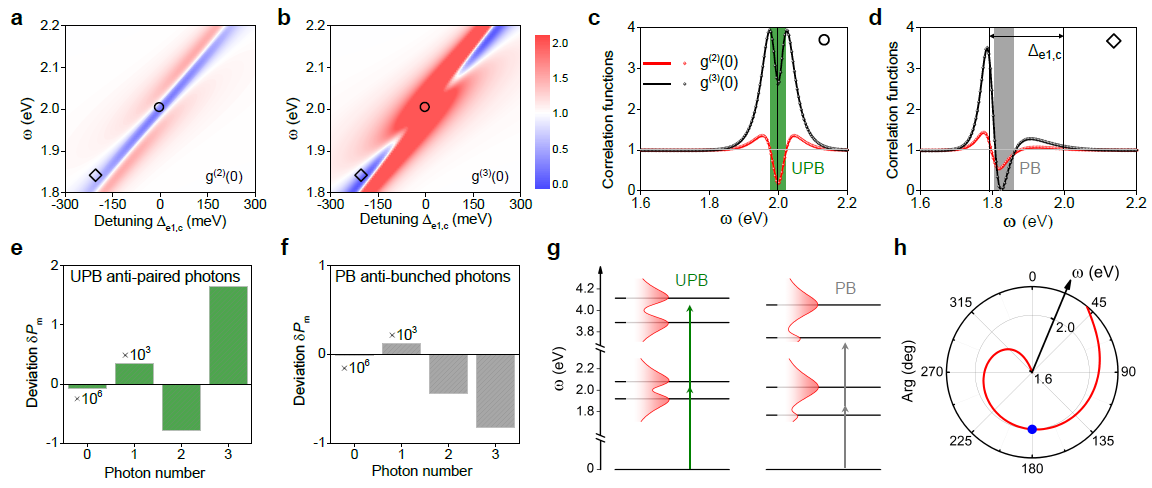}
\caption{
\textbf{Plexcitonic systems with indistinguishable emitters.}
(a)-(b) The effect of the detuning between the emitter and the cavity $\Delta_{\textrm{e1,c}}$ on the correlation functions
at zero delay $g^{(2)}(0)$ and $g^{(3)}(0)$.
Symbol ``circle" denotes a resonant system with UPB effect.
Symbol ``diamond" denotes an off-resonant system with PB effect.
(c)-(d) The two representative plexcitonic systems denoted by circle and diamond:
(c) anti-paired photon source based on UPB with $\Delta_{\textrm{e1,c}}=$ 0 and
(d) anti-bunched photon source based on PB with $\Delta_{\textrm{e1,c}}=-$ 205 meV.
The lines and the dots are the simulation results from the Lindblad master equation and the analytic solutions from the equations-of-motion method, respectively.
(e)-(f) Photon statistics. The suppressed two-photon emission probability and enhanced three-photon emission probability (${\delta}P_{2}<0,{\delta}P_{3}>0$) in (e) and the suppression of both two-photon and three-photon emission probabilities (${\delta}P_{2}<0,{\delta}P_{3}<0$) in (f) indicate the emergence of UPB and PB for the plexcitonic systems shown in (c) and (d), respectively.
(g) Calculated energy level diagrams, overlapped with the extinction spectra representing the density of states, for the plexcitonic systems in (c) and (d).
(h) The phase difference $\delta{\theta}$ between the interference pathways for the UPB system in (c).
In present study, $\omega_{\textrm{c}}=$ 2 eV, $\kappa=$ 350 meV,  $\gamma_{\textrm{e1}}=$ 80 meV, and $g_\textrm{e1}$ = 80 meV are fixed; $\omega_{\textrm{e1}}$ is the variable.
}
\label{fig2}
\end{figure}

We start our discussions with a plexcitonic system with indistinguishable emitters (notated as e$_1$), meaning that any additional emitters are identical.
In our study, the plasmonic cavity ($\omega_{\textrm{c}}=$ 2 eV, $\kappa=$ 350 meV) and the decay rate of emitter $\gamma_{\textrm{e1}}=$ 80 meV are fixed unless otherwise stated.
In general, we can set the state of this cavity-e$_1$ plexcitonic system by adjusting the detuning $\Delta_{\textrm{e1,c}}=\omega_{\textrm{e1}}-\omega_\textrm{c}$ or the coupling rate $g_\textrm{e1}$ between the plasmonic cavity and the emitter. As an example, only the effect of energy detuning $\Delta_{\textrm{e1,c}}$ will be analyzed here.
Figures \ref{fig2}a-\ref{fig2}b present $g^{(2)}(0)$ and $g^{(3)}(0)$ for plexcitonic systems bearing the same coupling rate of $g_\textrm{e1}=80$ meV but various detunings $\Delta_{\textrm{e1,c}}$.
Areas of both $g^{(2)}(0)<1$ and $g^{(3)}(0)<1$ indicate the conventional PB, and areas of $g^{(2)}(0)<1$ but $g^{(3)}(0)>1$ define UPB. Interestingly, photon blockade is always observed in a narrow blue/red region when the probing frequency $\omega$ is close to the frequency of emitter, because the nonlinearity of our plexcitonic system arises from the quantum emitter (a two-level system). Otherwise, when $\omega$ is far away from the frequency of emitter, only the coherent light can be observed, $i.e.$, the white-colour regions with $g^{(2)}(0)=1$ and $g^{(3)}(0)=1$.

Typically, we have resonant ($\Delta_{\textrm{e1,c}}=0$) and off-resonant ($\Delta_{\textrm{e1,c}} \neq 0$) systems.
Among all the systems bearing the same coupling rate of $g_\textrm{e1}=80$ meV shown in Figs. \ref{fig2}a-\ref{fig2}b
, a resonant system (indicated by the circle symbol) manifests the feature of UPB ($i.e.$, $g^{(2)}(0)<1$ and $g^{(3)}(0)>1$), while an off-resonant system (indicated by the diamond symbol) with detuning $|\Delta_{\textrm{e1,c}}|>100$ meV  will enter into the PB regime ($i.e.$, $g^{(2)}(0)<1$ and $g^{(3)}(0)<1$).
These two representative systems are illustrated in Figs. \ref{fig2}c-\ref{fig2}d.
For the UPB anti-paired system shown in Fig. \ref{fig2}c, two-photon emission is suppressed ($g^{(2)}(0)<1$) whereas three or more photons emit cooperatively ($g^{(3)}(0)>1$), when driven by a laser of energy around $\omega=$ 2 eV.
The other system in Fig. \ref{fig2}d, with a detuning of $\Delta_{\textrm{e1,c}}=-205$ meV, will support PB at a different laser energy around $\omega=$ 1.82 eV,
where the photon is emitted
one after another antibunchingly with $g^{(2)}(0)<1$ and $g^{(3)}(0)<1$.
Besides PB, there is another interesting peak at 1.78 eV, accounting for the photon bunching regime \cite{berthel2015photophysics} for our multi-level plexcitonic system.
In a two-level system, the transition occurs from the excited state to the ground state, and only one photon is generated, which is reflected by $g^{(2)}(0)<1$ constantly in the spectrum. However, for a multi-level system such as our plexcitonic system, there are other emission paths from $N$-photon Fock states ($N > 1$) to the ground state. These $N$-photon processes could generate bunching photons and lead to $g^{(2)}(0)>1$ and $g^{(3)}(0)>1$.

To illustrate the suppression and enhancement of $m$-photon emission unambiguously, we also analyze the photon statistics (see Methods) and present the photon-number distribution $P=\{P_{m}|m=0,1,2,...\}$, which is directly related to the photon correlation function, in Figs. \ref{fig2}e and \ref{fig2}f, for the two representative plexcitonic systems in Figs. \ref{fig2}c and \ref{fig2}d.
In the context of a lossy plexcitonic system, when the population of $m$-photons $P_{m}$ is suppressed (or enhanced), $i.e.$, $\delta{P}_{m}<$ 0 (or $>$ 0), the $m$-th order correlation is simultaneously suppressed (or enhanced), $i.e.$, $g^{(m)}(0)<$ 1 (or $>$ 1). Here, $\delta{P}_{m}=(P_{m}-\mathcal{P}_{m})/\mathcal{P}_{m}$ represents the relative deviation of a given photon-number distribution from the corresponding Poissonian distribution $\mathcal{P}_{m}$ of the coherent light.
From the photon statistics of the first system shown in Fig. \ref{fig2}e, two-photon emission is suppressed while three-photon emission is enhanced, corresponding to the characteristics of the UPB photon source.
In contrast, for the other system in Fig. \ref{fig2}f, both two-photon and three-photon emissions are suppressed, manifesting the PB feature.
In both cases, a slightly enhanced single photon emission are observed.
It is worth highlighting that this analysis of photon statistics provides a possible experimental route via photon counting experiments \cite{PhysRevLett.118.133604,PhysRevLett.121.047401,PhysRevLett.99.126403,PhysRevLett.100.067402,PhysRevB.81.033307,Assmann297} to reconstruct the correlation functions for verifying our proposal.

To explain the underlying mechanism, we plot the energy level diagrams of the above two plexcitonic systems in Fig. \ref{fig2}g.
On top of it, we also plot the calculated extinction spectra \cite{PhysRevLett.118.237401} showing Rabi splitting to represent the energy level broadening effect due to the finite decay rate.
The energy level diagram is recovered when the decay rate is set to zero.
Therefore we use this extinction spectra to describe the number of states that are available to be occupied in the plexcitonic system around each energy level, similar to the concept of density of states (DOS) in condensed matter physics.
It is interesting to note that the large decay rate of plasmonic cavity $\kappa=$ 350 meV effectively creates more optical states near each energy level, even at the dip of the Rabi splitting.
For the resonant UPB system driven by the laser field $\omega=$ 2 eV (green arrows), the second photon is able to enter into the second energy band and populate the energy states with certain probability. The second photons excited via distinct pathways interfere destructively, leading to the suppression of two-photon emission (see Fig. \ref{fig2}h below).
On the other hand, for the PB off-resonant system, the energy levels are shifted downward, accompanied by the asymmetric extinction spectra. When driven by a laser with $\omega=$ 1.82 eV (grey arrows), it is clear that the second photon is no longer able to enter into the second energy band, leading to PB.

Our numerical calculations can be analytically reproduced by the equations-of-motion method (see Supporting Information).
Briefly, we introduce anti-Hermitian terms into the Hamiltonian to describe the dissipations of the cavity and emitter \cite{Molmer:93}, then truncate the Hilbert space up to 3-plexciton states to approximately calculate the correlations at zero delay $g^{(2)}(0)$ and $g^{(3)}(0)$. The probability amplitudes of the 0-, 1-, 2- and 3-plexciton states are defined as $\{c_{1}\}$,
$\{c_{2}$, $c_{3}\}$, $\{c_{4}$, $c_{5}\}$, and
$\{c_{6}$, $c_{7}\}$, which can be derived analytically. Accordingly, $g^{(2)}(0)$ and $g^{(3)}(0)$ can be approximated as
$g^{(2)}(0)\approx\frac{2|c_{5}|^2}{|c_{2}|^4}$ and
$g^{(3)}(0)\approx\frac{6|c_{7}|^2}{|c_{2}|^6}$.
As shown in Figs. \ref{fig2}c-\ref{fig2}d, these analytic solutions are in excellent agreement with those obtained from the Lindblad master equation.

From the physical point of view, the suppression of two-photon emission in UPB is achieved by destructive interference. For example, it is straightforward to interpret the two-photon population  $|c_{5}|^2=|A_{52}c_{2}+A_{53}c_{3}|^2$ as the interference of two leading-order pathways: ($\#$1)
$|{0,0}\rangle\rightarrow|{1,0}\rangle\rightarrow|{2,0}\rangle$ and ($\#$2) $|{0,0}\rangle\rightarrow|{1,0}\rangle\rightarrow|{0,1}\rangle\rightarrow|{1,1}\rangle\rightarrow|{2,0}\rangle$ as shown in Fig. \ref{fig1}b.
The phase difference between the probability amplitudes for these two pathways $\delta\theta=\text{Arg}(A_{52}c_{2})-\text{Arg}(A_{53}c_{3})$ is plotted as a function of driven laser frequency $\omega$ in Fig. \ref{fig2}h. At $\omega=$ 2 eV where $g^{(2)}(0)$ reaches its minimum (Fig. \ref{fig2}c), the two pathways are out of phase ($\delta\theta=\pi$), as indicated by the blue dot. This is the evidence of destructive interference.

\subsection*{Plexcitonic systems with distinguishable emitters}

\begin{figure}[!ht]
\centering
\includegraphics[scale=0.75]{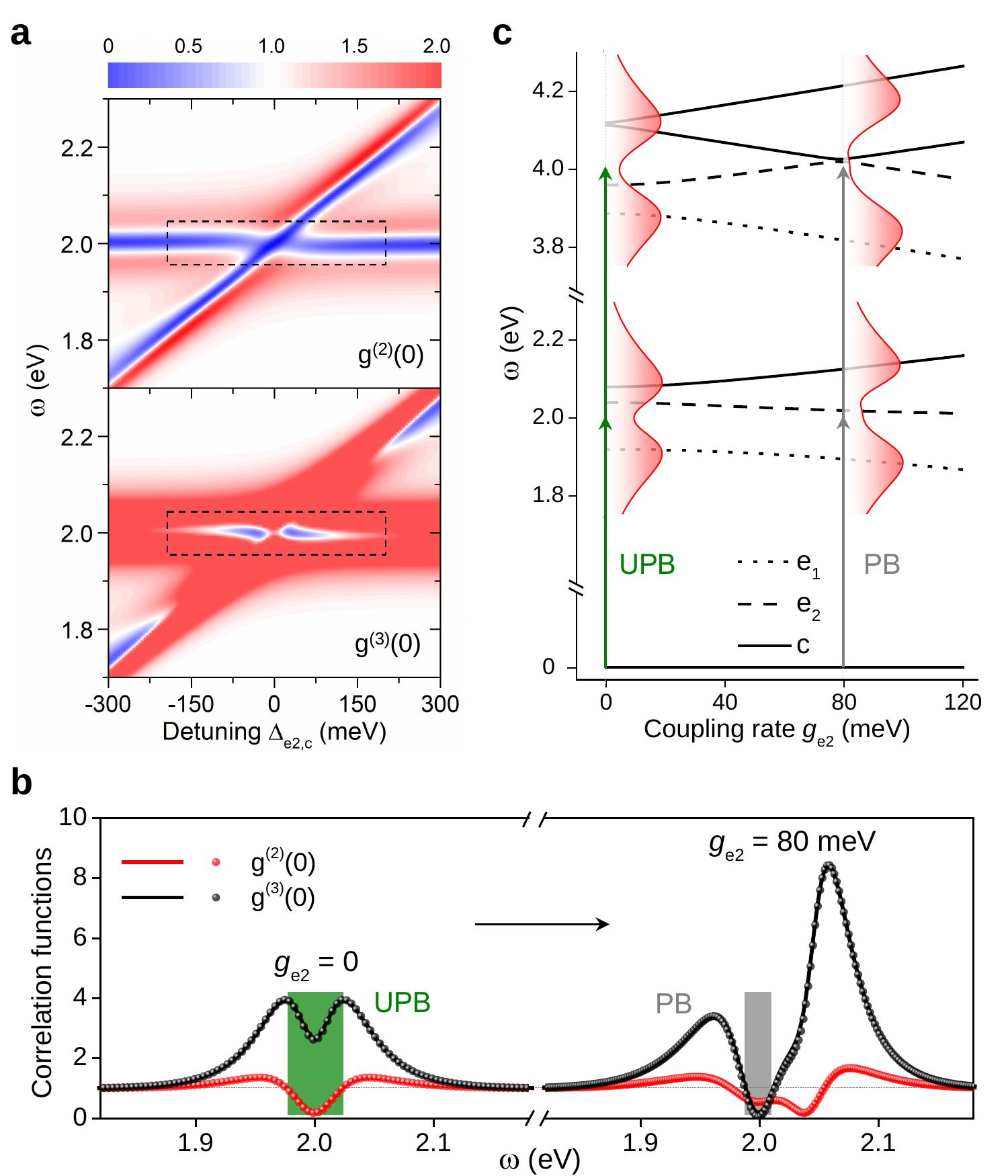}\\
\caption{
\textbf{Plexcitonic systems with distinguishable emitters.}
(a) The effect of the detuning between the second emitter and the cavity $\Delta_{\textrm{e2,c}}$ on the correlation functions  $g^{(2)}(0)$ and $g^{(3)}(0)$ for a fixed coupling rate  $g_{\textrm{e2}}=$ 80 meV. Dashed rectangles highlight the PB region.
(b) Case study of reconfiguration from UPB anti-paired to PB anti-bunched quantum light source by a second emitter e$_2$ slightly detuned from the cavity-e$_1$ system with $\Delta_{\textrm{e2,c}}=$ 40 meV. The lines and the dots are the simulation results from the Lindblad master equation and the analytic solutions from the equations-of-motion method, respectively.
(c) Calculated energy level diagrams as a function of the coupling rate between e$_2$ and the cavity $g_{\textrm{e2}}$, with fixed $\Delta_{\textrm{e2,c}}=$ 40 meV. Here, the extinction spectra for $g_{\textrm{e2}}=$ 0 and $g_{\textrm{e2}}=$ 80 meV are provided to indicate the DOS of these two plexcitonic systems.
In present study, a resonant cavity-e$_1$ system is employed with $\omega_{\textrm{c}}=\omega_{\textrm{e1}}=$ 2 eV, $\kappa=$ 350 meV, $\gamma_{\textrm{e1}}=$ 80 meV, and $g_\textrm{e1}=$ 80 meV.
The second emitter has fixed $\gamma_{\textrm{e2}}=$ 60 meV, whereas $\omega_{\textrm{e2}}$ and $g_{\textrm{e2}}$ are the variables.
}
\label{fig3}
\end{figure}

Once we have demonstrated that a quantum plexcitonic system with indistinguishable emitters could support either PB or UPB effect, we will show how the system can become actively reconfigurable between the two effects by introducing a second type of emitter e$_2$ into the picture. This also describes most plexcitonic systems with distinguishable emitters, as the subwavelength field localizations naturally make the quantum emitters spatially distinguishable.
In Fig. \ref{fig3}, we show a proof-of-concept study on how the second emitter e$_2$ modifies the emission property of the original cavity-e$_1$ plexcitonic system.
In the present study, the cavity-e$_1$ system is initially set into UPB state based on the resonant system ($\omega_{\textrm{c}}=\omega_{\textrm{e1}}=2$ eV) in Fig. \ref{fig2}c.
The second emitter e$_2$ has a fixed decay rate of $\gamma_{\textrm{e2}}=$ 60 meV, whereas $g_{\textrm{e2}}$ and $\omega_{\textrm{e2}}$ are the variables.
Figure \ref{fig3}a maps out $g^{(2)}(0)$ and $g^{(3)}(0)$ as a function of the detuning between e$_2$ and the cavity $\Delta_{\textrm{e2,c}}=\omega_{\textrm{e2}}-\omega_{\text{c}}$, when e$_2$ is coupled to the cavity with $g_{\textrm{e2}}=80$ meV.
As previously discussed, the photon blockade regions only appear when the probing frequency $\omega$ is close to the frequencies of the two emitters, as the nonlinearity of the plexcitonic system arises from the quantum emitters.
With the addition of the e$_2$, it is observed that PB can now be realized in the near-resonant blue islands highlighted by the dashed rectangles.

To elaborate the reconfiguration, in Fig. \ref{fig3}b, we start from the anti-paired UPB ($g^{(2)}(0)<1$ but $g^{(3)}(0)>1$) light source with $g_{\textrm{e2}}=$ 0 in Fig. \ref{fig2}c, and show that it changes to an anti-bunched PB ($g^{(2)}(0)<1$ and $g^{(3)}(0)<1$) light source when e$_2$ with small detuning $\Delta_{\textrm{e2,c}}=$ 40 meV  couples to the cavity with $g_{\textrm{e2}}=$ 80 meV, driven by the same laser at $\omega=$ 2 eV. Similarly, we can also obtain analytic solutions for $g^{(2)}(0)$ and $g^{(3)}(0)$ for the plexcitonic system with distinguishable emitters ($i.e.$, cavity-e$_1$-e$_2$) by equation-of-motion method (Supporting Information).
The analytic solutions of $g^{(2)}(0)$ and $g^{(3)}(0)$ in Fig. \ref{fig3}b again agree well with those obtained from the Lindblad master equation.

The PB region is found to be critically dependent on the coupling rate $g_{\textrm{e2}}$ between the second emitter and the cavity. To provide a simple physical picture, we plot the energy levels of the cavity-e$_1$-e$_2$ plexcitonic system as a function of $g_{\textrm{e2}}$ in Fig. \ref{fig3}c. With the participation of e$_2$, three (or four) energy levels appear for the first (or second) energy band.
For the first energy band around 2 eV, the second emitter e$_2$ (dashed line) will increase the energy level splittings of the cavity-e$_1$ system (solid and dotted lines), when the perturbation becomes stronger, $i.e.$, $g_{\textrm{e2}}$ is larger.
For the second energy band around 4 eV, in addition to the increased cavity-e$_1$ energy splitting, e$_2$ (dashed line) starts to strongly interact with the cavity (solid line), revealing a clear anti-crossing feature near $g_{\textrm{e2}}=$ 80 meV.
This results in a minimum DOS, blocking the absorption of second photon for the cavity-e$_1$-e$_2$ PB system, as indicated by the grey arrows.
In contrast, for the UPB system with indistinguishable emitters at $g_{\textrm{e2}}=$ 0 (green arrows), both the first and second photons occupy some optical energy states with much higher DOS.

The study above explains the mechanism to reconfigure a photon source between UPB and PB states in a plexcitonic system, when the  indistinguishable quantum emitters become distinguishable during a typical reconfiguration process. It is based on a simplified quantum optics model without taking into consideration of any plasmonic cavity design. In the following, we will illustrate the reconfiguration process in a realistic plasmonic cavity.

\subsection*{Realistic plexcitonic systems}

\begin{figure}[!ht]
\centering
\includegraphics[scale=0.55]{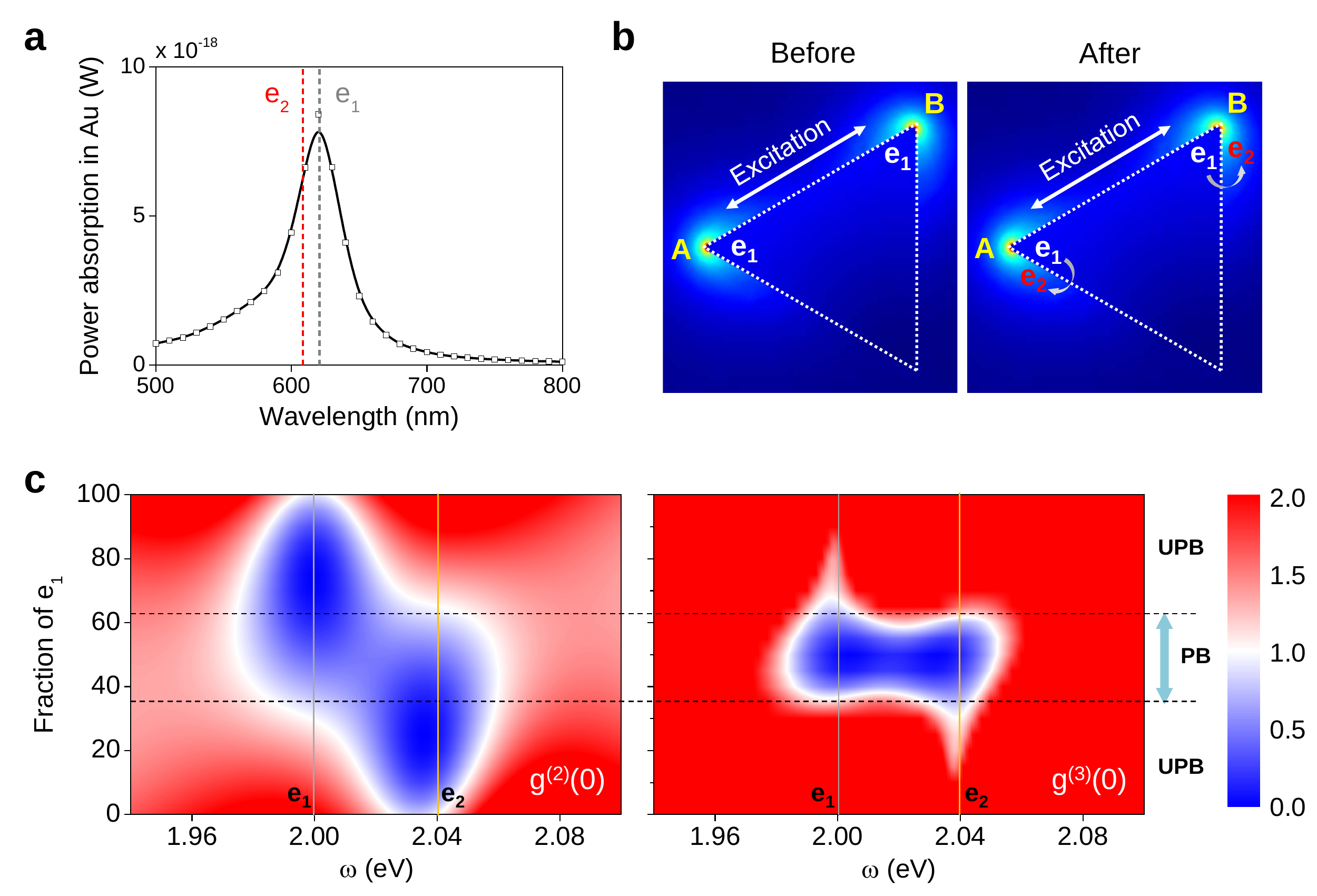}
\caption{
\textbf{Chemically reconfigurable single photon sources on nanoprisms.}
(a) The simulated absorption spectrum to identify the nanoprism plasmon mode, which is designed on resonance with the emitters e$_1$. Dashed grey and red lines represent the energies of the two types of emitters e$_1$ and e$_2$. (b) Schematics of the reconfigurable single photon source modulated via a chemical reaction. Before the chemical reaction (left), 100\% emitters e$_1$ are coupled to the nanoprism. After the chemical reaction (right), a portion of emitters e$_1$ is transformed to a new type of emitters e$_2$. (c) The effect of the fraction of e$_1$ on the correlation functions $g^{(2)}(0)$ and $g^{(3)}(0)$, with the PB region highlighted in light blue arrow.
}
\label{fig4}
\end{figure}

With rapid development in quantum plasmonics \cite{xu2018quantum,zhou2019quantum}, our proposal becomes experimentally feasible. In particular, the recently developed di-excitonic strong coupling system \cite{qian2018two,cuadra2018observation} would be an ideal experimental platform to realize the reconfigurable photon source.
Potential di-excitonic strong coupling systems include the monomer and aggregate forms of the same molecule, or photochromic spiropyran molecules \cite{schwartz2011reversible}, whose ratios could be chemically (or photo-chemically) and reversibly controlled.
This chemical scheme of reconfiguration typically changes the energy of emitter, making them distinguishable.
In Fig. \ref{fig4}, we elaborate the idea on an Au nanoprism with side length of 55 nm and a plasmon resonance at 2 eV or 620 nm. Initially, emitters e$_1$, which is on resonance with the nanoprism, are coated around the entire surface of the nanoprism.
When the chemical reaction starts, the emitters e$_1$ partially transform to a new type of emitters e$_2$ with the energy of $\omega_{\textrm{e2}}=$ 2.04 eV.

To generalize the problem, we treat the two types of emitters e$_1$ and e$_2$ at two different locations, A and B, as four distinguishable emitters. The Hamiltonian is written as:
\begin{equation}
\label{light_source}
\begin{split}
H&=\Delta_{\text{c}}a^{\dag}a+\Delta_{\textrm{A1}}\sigma_{\textrm{A1}}^{+}\sigma_{\textrm{A1}}^{-}+\Delta_{\textrm{B1}}\sigma_{\textrm{B1}}^{+}\sigma_{\textrm{B1}}^{-}+\Delta_{\textrm{A2}}\sigma_{\textrm{A2}}^{+}\sigma_{\textrm{A2}}^{-}+\Delta_{\textrm{B2}}\sigma_{\textrm{B2}}^{+}\sigma_{\textrm{B2}}^{-}+E_{l}(a+a^{\dag})\\
&+g_{\textrm{e1}}(a\sigma_{\textrm{A1}}^{+}+a^{\dag}\sigma_{\textrm{A1}}^{-})+g_{\textrm{e1}}(a\sigma_{\textrm{B1}}^{+}+a^{\dag}\sigma_{\textrm{B1}}^{-})
+g_{\textrm{e2}}(a\sigma_{\textrm{A2}}^{+}+a^{\dag}\sigma_{\textrm{A2}}^{-})+g_{\textrm{e2}}(a\sigma_{\textrm{B2}}^{+}+a^{\dag}\sigma_{\textrm{B2}}^{-}),\\
\end{split}
\end{equation}
where the frequencies are $\omega_{\text{c}}=2$ eV, $\omega_{\text{A1}}=\omega_{\text{B1}}=2$ eV representing e$_1$ at locations A and B, and $\omega_{\text{A2}}=\omega_{\text{B2}}=2.04$ eV representing e$_2$ at locations A and B.
Depending on the number ratio of e$_1$ and e$_2$, the coupling rates \cite{PhysRevLett.118.237401} are $g_{\textrm{e1}}=100\sqrt{f}$ and $g_{\textrm{e2}}=100\sqrt{1-f}$, respectively, with $0\le{f}\le1$ representing the fraction of e$_1$.
In this case, the coupling rate is independent of the location, due to the same electric fields at A and B.
Taking into account the dissipative terms, $\kappa=350$ meV, $\gamma_{\textrm{e1}}=\gamma_{\textrm{e2}}=60$ meV, the Lindblad master equation is then:
\begin{equation}
\label{mseq_light_source}
\begin{split}
\partial_{t}{\rho}&=i[\rho,H]+\frac{\kappa}{2}\mathcal{D}[a]\rho+\frac{\gamma_{\textrm{e1}}}{2}\mathcal{D}[\sigma_{\textrm{A1}}^{-}]\rho+\frac{\gamma_{\textrm{e1}}}{2}\mathcal{D}[\sigma_{\textrm{B1}}^{-}]\rho+\frac{\gamma_{\textrm{e2}}}{2}\mathcal{D}[\sigma_{\textrm{A2}}^{-}]\rho+\frac{\gamma_{\textrm{e2}}}{2}\mathcal{D}[\sigma_{\textrm{B2}}^{-}]\rho.\\
\end{split}
\end{equation}
As the fraction of e$_1$ begins dropping from 100$\%$ gradually to 0, we do observe the reconfiguration from  UPB to PB at the driving frequency of 2 eV (tuned to e$_1$), or from PB to UPB at the driving frequency of 2.04 eV (tuned to e$_2$), as shown by the calculated $g^{(2)}(0)$ and $g^{(3)}(0)$ in Fig. \ref{fig4}c.
If the driving frequency falls between 2 and 2.04 eV, we are able to see a reconfiguration from UPB to PB and then UPB again.
In short, a reconfigurable photon source between  anti-bunched (PB) and anti-paired (UPB) states can be realized by chemically controlling the ratio of e$_1$ and e$_2$. The effect originates from the different emitter energies of
$\omega_{\textrm{e1}}$ and $\omega_{\textrm{e2}}$, and the varied coupling rates of $g_{\textrm{e1}}$ and $g_{\textrm{e2}}$ along with the changed number ratio.
It should be noted that the coupling rate at the two extreme cases with all e$_1$ ($f=1$) or all e$_2$ ($f=0$) should be larger than 90 meV in order to observe this reconfiguration. The size of the PB region, $i.e.$, the blue area in $g^{(3)}(0)$, or the fraction range of e$_1$ to observe PB, is tunable by varying the coupling rate at the two extreme cases. With this number set to 100 meV, we observe the PB region with 40$-$60$\%$ of e$_1$ (see Fig. \ref{fig4}c). Increasing the coupling rate at the two extreme cases continuously reduces the area of PB region centralized around 50$\%$ of e$_1$, until it disappears at 170 meV. 

\begin{figure}[!ht]
\centering
\includegraphics[scale=0.6]{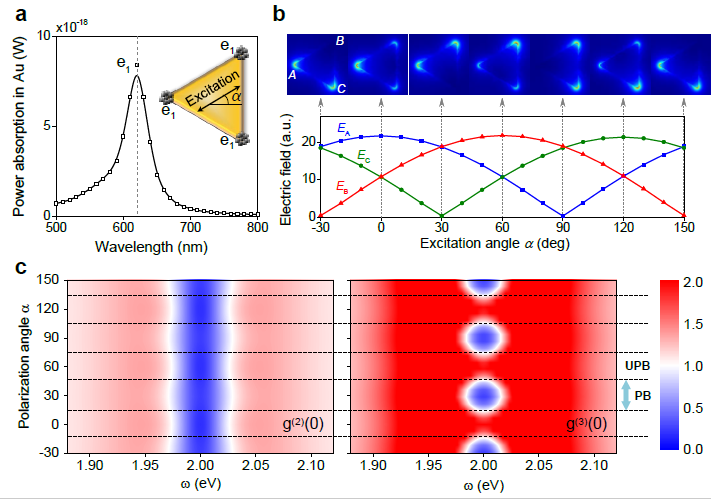}
\caption{
\textbf{Optically reconfigurable single photon sources on nanoprisms.}
(a) The simulated absorption spectrum to identify the nanoprism plasmon mode, which is designed on resonance with the emitters e$_1$. Inset: schematics of the reconfigurable single photon source modulated by the polarization angle $\alpha$ of the incident light, in which the coupling rate between the emitters and the nanoprism is strongly dependent on the local electric field. (b) The electric field distributions of the nanoprism plasmon mode (top) and the local electric fields at the three apexes (bottom) as the polarization angle $\alpha$ of incident light is rotated from $-30^\circ$ to $150^\circ$. (c) The effect of the incident polarization angle $\alpha$ on the correlation functions $g^{(2)}(0)$ and $g^{(3)}(0)$, with the PB region highlighted in light blue arrow.
}
\label{fig5}
\end{figure}

While the chemical scheme directly changes the properties of emitter, we could externally rotate the polarization angle $\alpha$ of the incident light to tune the coupling rate of emitters, as illustrated in Fig. \ref{fig5}a.
Starting with a similar nanoprism-e$_1$ resonant plexcitonic system, the emitters coated around the entire surface of the nanoprism are now distinguished by their locations, where we only mark down the three apexes with huge plasmonic field enhancements: A, B, C. As the polarization angle $\alpha$ rotates from $-30^\circ$ to $150^\circ$, the electric fields at the three locations change accordingly, as vividly shown in Fig. \ref{fig5}b. The electric fields have a direct impact on the local coupling rates \cite{PhysRevLett.118.237401}, which can be mathematically approximated as: $g_{\textrm{A}}=85|\cos{\alpha}|$, $g_{\textrm{B}}=85|\cos{(\alpha-60^\circ)}|$ and $g_{\textrm{C}}=85|\cos{(\alpha+60^\circ)}|$, following the same trend of $E_{\textrm{A}}$, $E_{\textrm{B}}$ and $E_{\textrm{C}}$ in Fig. \ref{fig5}b. These location-dependent coupling rates should be taken into the Hamiltonian:
\begin{equation}
\label{light_source}
\begin{split}
H&=\Delta_{\text{c}}a^{\dag}a+\Delta_{\textrm{A1}}\sigma_{\textrm{A1}}^{+}\sigma_{\textrm{A1}}^{-}+\Delta_{\textrm{B1}}\sigma_{\textrm{B1}}^{+}\sigma_{\textrm{B1}}^{-}+\Delta_{\textrm{C1}}\sigma_{\textrm{C1}}^{+}\sigma_{\textrm{C1}}^{-}+E_{l}(a+a^{\dag})\\
&+g_{\textrm{A}}(a\sigma_{\textrm{A1}}^{+}+a^{\dag}\sigma_{\textrm{A1}}^{-})+g_{\textrm{B}}(a\sigma_{\textrm{B1}}^{+}+a^{\dag}\sigma_{\textrm{B1}}^{-})
+g_{\textrm{C}}(a\sigma_{\textrm{C1}}^{+}+a^{\dag}\sigma_{\textrm{C1}}^{-}),\\
\end{split}
\end{equation}
where the frequencies are $\omega_{\text{c}}=\omega_{\text{A1}}=\omega_{\text{B1}}=\omega_{\text{C1}}=2$ eV, representing the same type of e$_1$ sitting at different locations.
These  same type of emitters e$_1$ become distinguishable due to the variations of the location-dependent coupling rates.
With $\kappa=350$ meV and $\gamma_{\textrm{e1}}=80$ meV, the Lindblad master equation is written as:
\begin{equation}
\label{mseq_light_source}
\begin{split}
\partial_{t}{\rho}&=i[\rho,H]+\frac{\kappa}{2}\mathcal{D}[a]\rho+\frac{\gamma_{\textrm{e1}}}{2}\mathcal{D}[\sigma_{\textrm{A1}}^{-}]\rho+\frac{\gamma_{\textrm{e1}}}{2}\mathcal{D}[\sigma_{\textrm{B1}}^{-}]\rho+\frac{\gamma_{\textrm{e1}}}{2}\mathcal{D}[\sigma_{\textrm{C1}}^{-}]\rho.\\
\end{split}
\end{equation}
The results of calculated
$g^{(2)}(0)$ and $g^{(3)}(0)$ are shown in Fig. \ref{fig5}c, where a repeating reconfiguration pattern of PB$-$UPB of cycle of $60^\circ$ are observed with the tuning polarization angle $\alpha$, when driven by a laser of 2 eV in resonance with the emitters.
The repeating cycle of $60^\circ$ in the correlation function of $g^{(3)}(0)$ in Fig. \ref{fig5}c follows the electric field distributions in Fig. \ref{fig5}b, where an equivalent field distribution appears every $60^\circ$.
Within each cycle of $60^\circ$, the modulation of $g^{(3)}(0)$ below and above one is necessary to realize the reconfiguration between UPB and PB, which can only be achieved with peak coupling rate between 80 and 100 meV. In this example, we are showing results of 85 meV.

\section*{Conclusions and outlook}
In conclusion, we have theoretically proposed a single photon source based on a quantum plexcitonic system with moderate nonlinearity. Either PB anti-bunched or UPB anti-paired photon source can be realized in a well-designed plexcitonic system with indistinguishable emitters. To turn the passive device active, a second type of emitters can be introduced to reconfigure the quantum photon source between the PB anti-bunched and UPB anti-paired states via tuning the energy band. In other words, this category of commonly seen plexcitonic systems with distinguishable emitters are particularly useful in realizing reconfigurable photon source. We demonstrate two realistic schemes of reconfiguration based on the same nanoprism plasmonic cavity, either chemically or optically. The chemical way directly changes the properties of the emitters, while the optical means externally varies the electric field distributions of the plasmonic cavity so as to change the coupling rate of the emitters at different locations. The latter, making emitters spatially distinguishable, is recognized as the key feature of plexcitonic systems.
Interesting further study includes investigating the collective excitation in the strong light-matter coupling system of multi-emitters and plasmonic cavities, which remarks the first step toward many-body phenomena such as optical superfluidity \cite{Greentree2006,Hartmann2006,PhysRevB.90.195112}.

From another perspective, the electromagnetic environment surrounding the plexcitonic system could influence the plasmon-emitter coupling. For example, in a vacuum environment, the radiative field originates from the near-field of the plasmon-emitter hybrid system and, as a consequence, there is no further coupling between the radiative field and the near-field. However, when an external microcavity is introduced into the system, the radiative field is now confined inside the microcavity, with its lifetime being related to the cavity quality factor. This idea leads to the concept of microcavity plasmonics \cite{xiao2012strongly,peng2017enhancing,xu2018quantum}, in which an embedding photonic microcavity is utilized to engineer the response of plasmonic nanostructures. In this way, both the high field concentration associated with localized surface plasmons and the low damping of cavity photons could be exploited to achieve a better performance.

\section{Methods}
\subsubsection{Quantum mechanical description based on Lindblad master equation}
A quantum emitter, such as an atom, a molecule, or a quantum dot, can be modeled by a two-level system with ground state $|g\rangle$ and excited state $|e\rangle$ separated by the transition frequency $\omega_{\text{e}1}$. The Hamiltonian of such an emitter is written as $\omega_{\text{e}1}\sigma_{\text{e}1}^{+}\sigma_{\text{e}1}^{-}$ with $\sigma_{\text{e}1}^{+}=|{e}\rangle\langle{g}|$ (or $\sigma_{\text{e}1}^{-}=|{g}\rangle\langle{e}|$) being the raising (or lowering) operator.
For classical light source with Poissonian and super-Poissonian distributions, the system can be described  by a semiclassical theory, where the plasmonic cavity is modeled as an electromagnetic wave and the emitter is modeled as a two-level system.
In contrast, sub-Poissonian light requires a quantization of the electromagnetic field to properly describe the particle nature of the light.
In this work, the single-mode plasmonic cavity is described by the standard Hamiltonian $\omega_{\text{c}}a^{\dag}a$, where $\omega_{\text{c}}$ is the cavity mode frequency, and $a$ is the photon annihilation operator obeying the bosonic commutation relation $[a,a^{\dag}]=1$. The interplay of emitter and cavity is mediated via the electric-dipole interaction with the coupling rate $g_{\text{e}1}$.
When $g_{\text{e}1}\ll\omega_{\text{c}},\omega_{\text{e}1}$, the quickly oscillating counter-rotating terms can be neglected by the rotating wave approximation. The resulting Hamiltonian is the well-known Jaynes-Cummings interaction, $g_{\text{e}1}(a\sigma_{\text{e}1}^{+}+a^{\dag}\sigma_{\text{e}1}^{-})$. To probe a nontrivial output of photons in the lossy environment, we drive the system using a weak coherent light, $E_{l}(ae^{i\omega{t}}+a^{\dag}e^{-i\omega{t}})$, with laser field strength $E_{l}$ and  laser frequency $\omega$.
Within the rotating frame \cite{Ste07,PhysRevA.100.053851},
our plasmonic cavity with single emitter can be described by the following Hamiltonian:
\begin{equation}
\label{light_source}
\begin{split}
H&=\Delta_{\text{c}}a^{\dag}a+\Delta_{\text{e}1}\sigma_{\text{e}1}^{+}\sigma_{\text{e}1}^{-}+g_{\text{e}1}(a\sigma_{\text{e}1}^{+}+a^{\dag}\sigma_{\text{e}1}^{-})+E_{l}(a+a^{\dag}),\\
\end{split}
\end{equation}
where $\Delta_{\text{c}}=\omega_{\text{c}}-\omega$ and $\Delta_{\text{e}1}=\omega_{\text{e}1}-\omega$ are the laser detunings for the cavity and emitter, respectively. Meanwhile, the plasmonic cavity and the emitter are exposed in a dissipative environment, which are described by the Lindblad master equation:
\begin{equation}
\label{mseq_light_source}
\begin{split}
\partial_{t}{\rho}&=i[\rho,H]+\frac{\kappa}{2}\mathcal{D}[a]\rho+\frac{\gamma_{\text{e}1}}{2}\mathcal{D}[\sigma_{\text{e}1}^{-}]\rho,\\
\end{split}
\end{equation}
where $\kappa$ and $\gamma_{\text{e}1}$ are the decay rates of the cavity and the emitter, respectively. The system is described by the density matrix $\rho$ and various dissipation channels for operator $\hat{o}$ are described by the Lindblad term $\mathcal{D}[\hat{o}]\rho=2\hat{o}\rho\hat{o}^{\dag}-\rho\hat{o}^{\dag}\hat{o}-\hat{o}^{\dag}\hat{o}\rho$.
The first term represents the coherent evolution of the system; the second and third terms correspond to the dissipations from the cavity and the emitter to the environment, respectively.

For the system with two emitters, the additional emitter can also be modeled as a two-level excitonic system $\omega_{\text{e}2}\sigma_{\text{e}2}^{+}\sigma_{\text{e}2}^{-}$, which is coupled to the plasmonic mode by the Jaynes-Cummings interaction, $g_{\text{e}2}(a\sigma_{\text{e}2}^{+}+a^{\dag}\sigma_{\text{e}2}^{-})$. Similarly, the spontaneous emission rate $\gamma_{\text{e}2}$ of the second emitter is taken into account by the Lindblad term $\frac{\gamma_{\text{e}2}}{2}\mathcal{D}[\sigma_{\text{e}2}^{-}]\rho$.

\subsubsection{Photon statistics}
Photon counting experiments\cite{PhysRevLett.121.047401,PhysRevLett.118.133604,PhysRevLett.99.126403,PhysRevLett.100.067402,PhysRevB.81.033307,Assmann297} allow us to reconstruct the photon-number distribution $P=\{P_{m}|m=0,1,2,...\}$ and to evaluate any orders of the correlation function at zero delay via the relationship:
\begin{equation}
\label{g_n}
\begin{split}
g^{(n)}(0)=\sum_{m=n}^{\infty}\frac{m!}{(m-n)!}\frac{P_{m}}{\langle{\hat{N}}\rangle^{n}},
\end{split}
\end{equation}
where $g^{(n)}(0)$ is the $n$-th order correlation, $P_{m}$ is the probability to measure $m$ photons, and $\langle{\hat{N}}\rangle=\sum_{m}mP_{m}$ is the mean photon number of the cavity field where $\hat{N}=a^{\dag}a$ is the particle number operator. Principally, we are able to determine the correlations up to the order of the highest measured photon number.
Thus the correlations of cavity field, such as $g^{(2)}(0)$ and $g^{(3)}(0)$ used in this work, can be reconstructed from the photon-number distribution.

Once the photon counting experiments produce the photon number distribution, the intrinsic statistical nature of photons in a light source could be analyzed. Three regimes of statistical distributions can be classified according to the properties of the light source: Poissonian, super-Poissonian, and sub-Poissonian, which correspond to the coherent light, classical light, and quantum light. To experimentally reveal the photon blockade effects, we can compare the photon-number distribution of the cavity field with the Poissonian distribution of the coherent light. For the state of coherent light $|\alpha\rangle=\sum_{m=0}^{\infty}\frac{\alpha^{m}}{\sqrt{m!}}e^{-\frac{|\alpha|^2}{2}}|m\rangle$, by projecting the coherent state onto the Fock state $|m\rangle$, the probability to find $m$ photons is exactly the Poissonian distribution:
\begin{equation}
\label{Poissonian}
\begin{split}
\mathcal{P}_{m}=\frac{|\alpha|^{2m}}{m!}e^{-|\alpha|^2}=\frac{\langle\hat{N}\rangle^{m}}{m!}e^{-\langle\hat{N}\rangle},
\end{split}
\end{equation}
which is a distinct feature of the coherent light source. For a cavity field characterized by the reduced steady density matrix $\rho^{\text{c,steady}}=\text{Tr}_{\text{e}}\rho^{\text{steady}}$, the probability to find $m$ photons is $P_{m}=\langle{m}|\rho^{\text{c,steady}}|{m}\rangle$. Here, $\text{Tr}_{\text{e}}$ is the partial trace over the sub-spaces of emitters. We thus arrive at a measure to show the relative deviation of a given photon-number distribution from the corresponding Poissonian distribution\cite{PhysRevLett.118.133604,PhysRevLett.121.153601}:
\begin{equation}
\label{derivative}
\begin{split}
\delta{P}_{m}=\frac{P_{m}-\mathcal{P}_{m}}{\mathcal{P}_{m}}.
\end{split}
\end{equation}
For the lossy plasmonic nanostructure in the weak-driving regime, the photon-number distribution fulfills the condition $P_{m} \gg P_{m+1}$ and the mean photon number in the steady state is nearly zero, $\langle{N}\rangle\approx0$.
Thus the first term ($i.e.$, $m=n$) is the dominant term in the $n$-th order correlation in Eq. (\ref{g_n}). By replacing the index $n$ by $m$, we have the $m$-th order correlation:
\begin{equation}
\begin{split}
g^{(m)}(0)\approx \frac{m!P_{m}}{\langle{N}\rangle^m},
\end{split}
\end{equation}
and the Poissonian distribution can be approximated as:
\begin{equation}
\begin{split}
\mathcal{P}_{m}\approx
\frac{\langle\hat{N}\rangle^{m}}{m!}.
\end{split}
\end{equation}
In short, when the population of $m$ photons $P_{m}$ is suppressed (or enhanced), $i.e.$, $\delta{P}_{m}<$ 0 (or $>$ 0), the $m$-th order correlation is simultaneously suppressed (enhanced), $i.e.$, $g^{(m)}(0)<$ 1 (or $>$ 1). Therefore, the photon-number distribution is directly related to the photon correlation function in the lossy plexcitonic system.

\begin{acknowledgement}
Institute of High Performance Computing (IHPC) acknowledges financial support from the National Research Foundation Singapore NRF2017-NRF-NSFC002-015, NRF2016-NRF-ANR002; and A*STAR SERC A1685b0005. W.-L. Yang  acknowledges financial supports from the Youth Innovation Promotion Association (CAS No. 2016299).
\end{acknowledgement}


\newcommand{\beginsupplement}{%
        \setcounter{table}{0}
        \renewcommand{\thetable}{S\arabic{table}}%
        \setcounter{figure}{0}
        \renewcommand{\thefigure}{S\arabic{figure}}%
        \setcounter{equation}{0}
        \renewcommand{\theequation}{S\arabic{equation}}%
     }

\section{Supporting Information}
\beginsupplement

\section*{\large Equations-of-motion method to obtain the correlations at zero delay}

To explain the numerical results from Lindblad master equation and understand the quantum interference mechanism for UPB, we derive a set of analytic solutions based on equations-of-motion method.
For the single-emitter system, we introduce anti-Hermitian terms to the Hamiltonian in Eq. (8) in the main text to describe the dissipation of the cavity and the emitter according to the quantum trajectory method\cite{Molmer:93}. The effective non-Hermitian Hamiltonian is thus given by:
\begin{equation}
\begin{split}
H_{\text{eff}}&=\Delta_{\text{c}}^{'}a^{\dag}a+\Delta_{\text{e}1}^{'}\sigma_{\text{e}1}^{+}\sigma_{\text{e}1}^{-}+g_{\text{e}1}(a\sigma_{\text{e}1}^{+}+a^{\dag}\sigma_{\text{e}1}^{-})+E_{l}(a+a^{\dag}),\\
\end{split}
\end{equation}
where $\Delta_{\text{c}}^{'}=\Delta_{\text{c}}-i\kappa/2$ and $\Delta_{\text{e}1}^{'}=\Delta_{\text{e}1}-i\gamma_{\text{e}1}/2$. To calculate the correlations at zero delay up to the third order, the Hilbert space needs to be truncated at least to the $3$-plexciton subspace. Therefore, the wave-function can be written as:
\begin{equation}
\label{psi}
\begin{split}
|\psi(t)\rangle&=c_{1}(t)|{0,0}\rangle+c_{2}(t)|{1,0}\rangle+c_{3}(t)|{0,1}\rangle+c_{4}(t)|{1,1}\rangle+c_{5}(t)|{2,0}\rangle+c_{6}(t)|{2,1}\rangle+c_{7}(t)|{3,0}\rangle.\\
\end{split}
\end{equation}
Substituting the wave-function in Eq. (\ref{psi}) into the Schr\"{o}dinger equation, $i\partial_{t}|\psi(t)\rangle=H_{\text{eff}}|\psi(t)\rangle$, we have\cite{Deng:17,PhysRevLett.121.153601}:
\begin{equation}
\label{eom2}
\begin{split}
i\dot{c_{1}}(t)&=E_{l}c_{2}(t),\\
i\dot{c_{2}}(t)&=\Delta_{\text{c}}^{'}c_{2}(t)+g_{\text{e}1}c_{3}(t)+E_{l}c_{1}(t)+\sqrt{2}E_{l}c_{5}(t),\\
i\dot{c_{3}}(t)&=\Delta_{\text{e}1}^{'}c_{3}(t)+g_{\text{e}1}c_{2}(t)+E_{l}c_{4}(t),\\
i\dot{c_{4}}(t)&=(\Delta_{\text{c}}^{'}+\Delta_{\text{e}1}^{'})c_{4}(t)+\sqrt{2}g_{\text{e}1}c_{5}(t)+E_{l}c_{3}(t)+\sqrt{2}E_{l}c_{6}(t),\\
i\dot{c_{5}}(t)&=2\Delta_{\text{c}}^{'}c_{5}(t)+\sqrt{2}g_{\text{e}1}c_{4}(t)+\sqrt{2}E_{l}c_{2}(t)+\sqrt{3}E_{l}c_{7}(t),\\
i\dot{c_{6}}(t)&=(2\Delta_{\text{c}}^{'}+\Delta_{\text{e}1}^{'})c_{6}(t)+\sqrt{3}g_{\text{e}1}c_{7}(t)+\sqrt{2}E_{l}c_{4}(t),\\
i\dot{c_{7}}(t)&=3\Delta_{\text{c}}^{'}c_{7}(t)+\sqrt{3}g_{\text{e}1}c_{6}(t)+\sqrt{3}E_{l}c_{5}(t).\\ \\
\end{split}
\end{equation}
Under weak-driving condition, the steady probability amplitudes should satisfy the relations:
\begin{equation}
\begin{split}
\{c_{1}\}\gg\{c_{2},c_{3}\}\gg\{c_{4},c_{5}\}\gg\{c_{6},c_{7}\}.\\
\end{split}
\end{equation}
We can approximately solve Eq. (\ref{eom2}) using a perturbation method by discarding higher-order terms in each equation for lower-order variables. For the probability amplitude $\{c_{1}\}$ of the $0$-plexciton state, $i\dot{c_{1}}\approx0$ thus we have $c_{1}\approx1$. This is reasonable under the weak-driving and large decay conditions, \textit{i.e.}, photons are inclined to decay into the ground state.

For the steady probability amplitudes $\{c_{2},c_{3}\}$ of the $1$-plexciton states, we can make a three-state truncation of the Hilbert space. Keeping only the ground state $|{0,0}\rangle$ and the excited states $\{|{1,0}\rangle,|{0,1}\rangle\}$, and setting the right-hand side of Eq. (\ref{eom2}) to zero, we have:
\begin{equation}
\begin{split}
&\Delta_{\text{c}}^{'}c_{2}+g_{\text{e}1}c_{3}+E_{l}=0,\\
&\Delta_{\text{e}1}^{'}c_{3}+g_{\text{e}1}c_{2}=0.\\
\end{split}
\end{equation}
Thus the probability amplitudes for the $1$-plexciton states are:
\begin{equation}
\label{c23}
\begin{split}
c_{2}&=E_{l}\frac{-\Delta_{\text{e}1}^{'}}{\Delta_{\text{c}}^{'}\Delta_{\text{e}1}^{'}-g_{\text{e}1}^2},\hspace{0.2cm}c_{3}=E_{l}\frac{g_{\text{e}1}}{\Delta_{\text{c}}^{'}\Delta_{\text{e}1}^{'}-g_{\text{e}1}^2}.\\
\end{split}
\end{equation}
Similarly, for the steady probability amplitudes $\{c_{4},c_{5}\}$ of the $2$-plexciton states, we can make a five-state truncation of the Hilbert space, then we have:
\begin{equation}
\begin{split}
&(\Delta_{\text{c}}^{'}+\Delta_{\text{e}1}^{'})c_{4}+\sqrt{2}g_{\text{e}1}c_{5}+E_{l}c_{3}=0,\\
&2\Delta_{\text{c}}^{'}c_{5}+\sqrt{2}g_{\text{e}1}c_{4}+\sqrt{2}E_{l}c_{2}=0,\\
\end{split}
\end{equation}
and therefore,
\begin{equation}
\label{c45}
\begin{split}
c_{4}&=A_{42}c_{2}+A_{43}c_{3},\hspace{0.2cm}c_{5}=A_{52}c_{2}+A_{53}c_{3},\\
A_{42}&=E_{l}\frac{g_{\text{e}1}}{\Delta_{\text{c}}^{'}(\Delta_{\text{c}}^{'}+\Delta_{\text{e}1}^{'})-g_{\text{e}1}^2},\hspace{0.2cm}A_{43}=E_{l}\frac{-\Delta_{\text{c}}^{'}}{\Delta_{\text{c}}^{'}(\Delta_{\text{c}}^{'}+\Delta_{\text{e}1}^{'})-g_{\text{e}1}^2},\\
A_{52}&=\frac{E_{l}}{\sqrt{2}}\frac{-(\Delta_{\text{c}}^{'}+\Delta_{\text{e}1}^{'})}{\Delta_{\text{c}}^{'}(\Delta_{\text{c}}^{'}+\Delta_{\text{e}1}^{'})-g_{\text{e}1}^2},\hspace{0.2cm}A_{53}=\frac{E_{l}}{\sqrt{2}}\frac{g_{\text{e}1}}{\Delta_{\text{c}}^{'}(\Delta_{\text{c}}^{'}+\Delta_{\text{e}1}^{'})-g_{\text{e}1}^2}.\\
\end{split}
\end{equation}
Thus the second-order correlation at zero delay is given by:
\begin{equation}
\label{g20_eom}
\begin{split}
g^{(2)}(0)=\frac{\langle{a^{\dag 2}a^{2}}\rangle}{\langle{a^{\dag}a}\rangle^{2}}=\frac{2|c_{5}|^2}{(|c_{2}|^2+|c_{4}|^2+2|c_{5}|^2)^2}\approx\frac{2|c_{5}|^2}{|c_{2}|^4}.\\
\end{split}
\end{equation}
Similarly, for the steady probability amplitudes $\{c_{6},c_{7}\}$ of the $3$-plexciton states,
\begin{equation}
\begin{split}
&(2\Delta_{\text{c}}^{'}+\Delta_{\text{e}1}^{'})c_{6}+\sqrt{3}g_{\text{e}1}c_{7}+\sqrt{2}E_{l}c_{4}=0,\\
&3\Delta_{\text{c}}^{'}c_{7}+\sqrt{3}g_{\text{e}1}c_{6}+\sqrt{3}E_{l}c_{5}=0,\\
\end{split}
\end{equation}
we have:
\begin{equation}
\label{c67}
\begin{split}
c_{6}=E_{l}\frac{-\sqrt{2}\Delta_{\text{c}}^{'}c_{4}+g_{\text{e}1}c_{5}}{\Delta_{\text{c}}^{'}(2\Delta_{\text{c}}^{'}+\Delta_{\text{e}1}^{'})-g_{\text{e}1}^2},\hspace{0.2cm}c_{7}=\frac{E_{l}}{\sqrt{3}}\frac{\sqrt{2}g_{\text{e}1}c_{4}-(2\Delta_{\text{c}}^{'}+\Delta_{\text{e}1}^{'})c_{5}}{\Delta_{\text{c}}^{'}(2\Delta_{\text{c}}^{'}+\Delta_{\text{e}1}^{'})-g_{\text{e}1}^2},\\
\end{split}
\end{equation}
and the third-order correlation at zero delay is:
\begin{equation}
\label{g30_eom}
\begin{split}
g^{(3)}(0)=\frac{\langle{a^{\dag 3}a^{3}}\rangle}{\langle{a^{\dag}a}\rangle^{3}}=\frac{6|c_{7}|^2}{(|c_{2}|^2+|c_{4}|^2+2|c_{5}|^2+2|c_{6}|^2+3|c_{7}|^2)^3}\approx\frac{6|c_{7}|^2}{|c_{2}|^6}.\\
\end{split}
\end{equation}

On the other hand, the similar derivation procedure can be applied to the system with two emitters. First of all, the effective non-Hermitian Hamiltonian is given by:
\begin{equation}
\begin{split}
H_{\text{eff}}&=\Delta_{\text{c}}^{'}a^{\dag}a+\Delta_{\text{e}1}^{'}\sigma_{\text{e}1}^{+}\sigma_{\text{e}1}^{-}+\Delta_{\text{e}2}^{'}\sigma_{\text{e}2}^{+}\sigma_{\text{e}2}^{-}+g_{\text{e}1}(a\sigma_{\text{e}1}^{+}+a^{\dag}\sigma_{\text{e}1}^{-})+g_{\text{e}2}(a\sigma_{\text{e}2}^{+}+a^{\dag}\sigma_{\text{e}2}^{-})+E_{l}(a+a^{\dag}),\\
\end{split}
\end{equation}
where $\Delta_{\text{e}2}^{'}=\Delta_{\text{e}2}-i\gamma_{\text{e}2}/2$. To calculate the correlations at zero delay up to third order, similarly, the Hilbert space needs to be truncated at least to the $3$-plexciton subspace. Therefore, the wave-function can be written as:
\begin{equation}
\begin{split}
|\psi(t)\rangle&=c_{1}(t)|{0,0,0}\rangle\\
&+c_{2}(t)|{1,0,0}\rangle+c_{3}(t)|{0,1,0}\rangle+c_{4}(t)|{0,0,1}\rangle\\
&+c_{5}(t)|{2,0,0}\rangle+c_{6}(t)|{1,1,0}\rangle+c_{7}(t)|{1,0,1}\rangle+c_{8}(t)|{0,1,1}\rangle\\
&+c_{9}(t)|{3,0,0}\rangle+c_{10}(t)|{2,1,0}\rangle+c_{11}(t)|{2,0,1}\rangle+c_{12}(t)|{1,1,1}\rangle.\\ \\
\end{split}
\end{equation}
Under weak-driving condition, the steady probability amplitudes satisfy the relations:
\begin{equation}
\begin{split}
\{c_{1}\}\gg\{c_{2},c_{3},c_{4}\}\gg\{c_{5},c_{6},c_{7},c_{8}\}\gg\{c_{9},c_{10},c_{11},c_{12}\}.\\
\end{split}
\end{equation}
Similar to the single-emitter system, the photons are inclined to decay into the ground state under the condition of weak-driving and large decay, which lead to $i\dot{c_{1}}\approx0$ and $c_{1}\approx1$. For the steady probability amplitudes $\{c_{2},c_{3},c_{4}\}$ of the $1$-plexciton states, we can truncate the Hilbert space up to the $1$-plexciton states, then the equations for the steady probability amplitudes are given by:
\begin{equation}
\begin{split}
&\Delta_{\text{c}}^{'}c_{2}+g_{\text{e}1}c_{3}+g_{\text{e}2}c_{4}+E_{l}=0,\\
&\Delta_{\text{e}1}^{'}c_{3}+g_{\text{e}1}c_{2}=0,\\
&\Delta_{\text{e}2}^{'}c_{4}+g_{\text{e}2}c_{2}=0.\\
\end{split}
\end{equation}
Thus the steady probability amplitudes for the $1$-plexciton states are:
\begin{equation}
\label{c234}
\begin{split}
c_{2}&=E_{l}\frac{-\Delta_{\text{e}1}^{'}\Delta_{\text{e}2}^{'}}{\Delta_{\text{c}}^{'}\Delta_{\text{e}1}^{'}\Delta_{\text{e}2}^{'}-g_{\text{e}1}^2\Delta_{\text{e}2}^{'}-g_{\text{e}2}^{2}\Delta_{\text{e}1}^{'}},\\
c_{3}&=E_{l}\frac{g_{\text{e}1}\Delta_{\text{e}2}^{'}}{\Delta_{\text{c}}^{'}\Delta_{\text{e}1}^{'}\Delta_{\text{e}2}^{'}-g_{\text{e}1}^2\Delta_{\text{e}2}^{'}-g_{\text{e}2}^{2}\Delta_{\text{e}1}^{'}},\\
c_{4}&=E_{l}\frac{g_{\text{e}2}\Delta_{\text{e}1}^{'}}{\Delta_{\text{c}}^{'}\Delta_{\text{e}1}^{'}\Delta_{\text{e}2}^{'}-g_{\text{e}1}^2\Delta_{\text{e}2}^{'}-g_{\text{e}2}^{2}\Delta_{\text{e}1}^{'}}.\\
\end{split}
\end{equation}
Similarly, for the steady probability amplitudes $\{c_{5},c_{6},c_{7},c_{8}\}$ of the $2$-plexciton states, we can truncate the Hilbert space up to the $2$-plexciton states, then the steady probability amplitudes can be obtained by solving the following equations: \begin{equation}
\label{c5678}
\begin{split}
&\left[\begin{array}{*{20}c}
{2\Delta_{\text{c}}^{'}} & {\sqrt{2}g_{\text{e}1}} & {\sqrt{2}g_{\text{e}2}} & {0}\\
{\sqrt{2}g_{\text{e}1}} & {\Delta_{\text{c}}^{'}+\Delta_{\text{e}1}^{'}} & {0} & {g_{\text{e}2}}\\
{\sqrt{2}g_{\text{e}2}} & {0} & {\Delta_{\text{c}}^{'}+\Delta_{\text{e}2}^{'}} & {g_{\text{e}1}}\\
{0} & {g_{\text{e}2}} & {g_{\text{e}1}} & {\Delta_{\text{e}1}^{'}+\Delta_{\text{e}2}^{'}}\\
\end{array}\right]\left[\begin{array}{*{20}c}
{c_{5}}\\
{c_{6}}\\
{c_{7}}\\
{c_{8}}\\
\end{array}\right]=-E_{l}\left[\begin{array}{*{20}c}
{\sqrt{2}c_{2}}\\
{c_{3}}\\
{c_{4}}\\
{0}\\
\end{array}\right].\\
\end{split}
\end{equation}
Therefore, the second-order correlation function at zero delay is given by:
\begin{equation}
\label{g20_eom2}
\begin{split}
g^{(2)}(0)=\frac{2|c_{5}|^2}{(|c_{2}|^2+2|c_{5}|^2+|c_{6}|^2+|c_{7}|^2)^2}\approx\frac{2|c_{5}|^2}{|c_{2}|^4}.\\
\end{split}
\end{equation}
Similarly, we can obtain the steady probability amplitudes $\{c_{9},c_{10},c_{11},c_{12}\}$ of the $3$-plexciton states by solving equations:
\begin{equation}
\label{c9101112}
\begin{split}
&\left[\begin{array}{*{20}c}
{3\Delta_{\text{c}}^{'}} & {\sqrt{3}g_{\text{e}1}} & {\sqrt{3}g_{\text{e}2}} & {0}\\
{\sqrt{3}g_{\text{e}1}} & {2\Delta_{\text{c}}^{'}+\Delta_{\text{e}1}^{'}} & {0} & {\sqrt{2}g_{\text{e}2}}\\
{\sqrt{3}g_{\text{e}2}} & {0} & {2\Delta_{\text{c}}^{'}+\Delta_{\text{e}2}^{'}} & {\sqrt{2}g_{\text{e}1}}\\
{0} & {\sqrt{2}g_{\text{e}2}} & {\sqrt{2}g_{\text{e}1}} & {\Delta_{\text{c}}^{'}+\Delta_{\text{e}1}^{'}+\Delta_{\text{e}2}^{'}}\\
\end{array}\right]\left[\begin{array}{*{20}c}
{c_{9}}\\
{c_{10}}\\
{c_{11}}\\
{c_{12}}\\
\end{array}\right]=-E_{l}\left[\begin{array}{*{20}c}
{\sqrt{3}c_{5}}\\
{\sqrt{2}c_{6}}\\
{\sqrt{2}c_{7}}\\
{c_{8}}\\
\end{array}\right].\\
\end{split}
\end{equation}
The third-order correlation function at zero delay is then given by:
\begin{equation}
\label{g30_eom2}
\begin{split}
g^{(3)}(0)=\frac{6|c_{9}|^2}{(|c_{2}|^2+2|c_{5}|^2+|c_{6}|^2+|c_{7}|^2+3|c_{9}|^2+2|c_{10}|^2+2|c_{11}|^2+|c_{12}|^2)^3}\approx\frac{6|c_{9}|^2}{|c_{2}|^6}.\\
\end{split}
\end{equation}

\bibliography{biblio}

\providecommand{\latin}[1]{#1}
\makeatletter
\providecommand{\doi}
  {\begingroup\let\do\@makeother\dospecials
  \catcode`\{=1 \catcode`\}=2 \doi@aux}
\providecommand{\doi@aux}[1]{\endgroup\texttt{#1}}
\makeatother
\providecommand*\mcitethebibliography{\thebibliography}
\csname @ifundefined\endcsname{endmcitethebibliography}
  {\let\endmcitethebibliography\endthebibliography}{}
\begin{mcitethebibliography}{57}
\providecommand*\natexlab[1]{#1}
\providecommand*\mciteSetBstSublistMode[1]{}
\providecommand*\mciteSetBstMaxWidthForm[2]{}
\providecommand*\mciteBstWouldAddEndPuncttrue
  {\def\EndOfBibitem{\unskip.}}
\providecommand*\mciteBstWouldAddEndPunctfalse
  {\let\EndOfBibitem\relax}
\providecommand*\mciteSetBstMidEndSepPunct[3]{}
\providecommand*\mciteSetBstSublistLabelBeginEnd[3]{}
\providecommand*\EndOfBibitem{}
\mciteSetBstSublistMode{f}
\mciteSetBstMaxWidthForm{subitem}{(\alph{mcitesubitemcount})}
\mciteSetBstSublistLabelBeginEnd
  {\mcitemaxwidthsubitemform\space}
  {\relax}
  {\relax}

\bibitem[Scarani \latin{et~al.}(2009)Scarani, Bechmann-Pasquinucci, Cerf,
  Du\ifmmode~\check{s}\else \v{s}\fi{}ek, L\"utkenhaus, and
  Peev]{RevModPhys.81.1301}
Scarani,~V.; Bechmann-Pasquinucci,~H.; Cerf,~N.~J.; Du\ifmmode~\check{s}\else
  \v{s}\fi{}ek,~M.; L\"utkenhaus,~N.; Peev,~M. The security of practical
  quantum key distribution. \emph{Rev. Mod. Phys.} \textbf{2009}, \emph{81},
  1301--1350\relax
\mciteBstWouldAddEndPuncttrue
\mciteSetBstMidEndSepPunct{\mcitedefaultmidpunct}
{\mcitedefaultendpunct}{\mcitedefaultseppunct}\relax
\EndOfBibitem
\bibitem[Herrero-Collantes and Garcia-Escartin(2017)Herrero-Collantes, and
  Garcia-Escartin]{RevModPhys.89.015004}
Herrero-Collantes,~M.; Garcia-Escartin,~J.~C. Quantum random number generators.
  \emph{Rev. Mod. Phys.} \textbf{2017}, \emph{89}, 015004\relax
\mciteBstWouldAddEndPuncttrue
\mciteSetBstMidEndSepPunct{\mcitedefaultmidpunct}
{\mcitedefaultendpunct}{\mcitedefaultseppunct}\relax
\EndOfBibitem
\bibitem[Pezz\`e \latin{et~al.}(2018)Pezz\`e, Smerzi, Oberthaler, Schmied, and
  Treutlein]{RevModPhys.90.035005}
Pezz\`e,~L.; Smerzi,~A.; Oberthaler,~M.~K.; Schmied,~R.; Treutlein,~P. Quantum
  metrology with nonclassical states of atomic ensembles. \emph{Rev. Mod.
  Phys.} \textbf{2018}, \emph{90}, 035005\relax
\mciteBstWouldAddEndPuncttrue
\mciteSetBstMidEndSepPunct{\mcitedefaultmidpunct}
{\mcitedefaultendpunct}{\mcitedefaultseppunct}\relax
\EndOfBibitem
\bibitem[Chunnilall \latin{et~al.}(2014)Chunnilall, Degiovanni, K\"{u}ck,
  M\"{u}ller, and Sinclair]{doi:10.1117/1.OE.53.8.081910}
Chunnilall,~C.~J.; Degiovanni,~I.~P.; K\"{u}ck,~S.; M\"{u}ller,~I.;
  Sinclair,~A.~G. Metrology of single-photon sources and detectors: a review.
  \emph{Optical Engineering} \textbf{2014}, \emph{53}, 081910\relax
\mciteBstWouldAddEndPuncttrue
\mciteSetBstMidEndSepPunct{\mcitedefaultmidpunct}
{\mcitedefaultendpunct}{\mcitedefaultseppunct}\relax
\EndOfBibitem
\bibitem[Dayan \latin{et~al.}(2008)Dayan, Parkins, Aoki, Ostby, Vahala, and
  Kimble]{Dayan1062}
Dayan,~B.; Parkins,~A.~S.; Aoki,~T.; Ostby,~E.~P.; Vahala,~K.~J.; Kimble,~H.~J.
  A Photon Turnstile Dynamically Regulated by One Atom. \emph{Science}
  \textbf{2008}, \emph{319}, 1062--1065\relax
\mciteBstWouldAddEndPuncttrue
\mciteSetBstMidEndSepPunct{\mcitedefaultmidpunct}
{\mcitedefaultendpunct}{\mcitedefaultseppunct}\relax
\EndOfBibitem
\bibitem[S{\'a}ez-Bl{\'a}zquez \latin{et~al.}(2017)S{\'a}ez-Bl{\'a}zquez,
  Feist, Fern{\'a}ndez-Dom{\'\i}nguez, and
  Garc{\'\i}a-Vidal]{saez2017enhancing}
S{\'a}ez-Bl{\'a}zquez,~R.; Feist,~J.; Fern{\'a}ndez-Dom{\'\i}nguez,~A.;
  Garc{\'\i}a-Vidal,~F. Enhancing photon correlations through plasmonic strong
  coupling. \emph{Optica} \textbf{2017}, \emph{4}, 1363--1367\relax
\mciteBstWouldAddEndPuncttrue
\mciteSetBstMidEndSepPunct{\mcitedefaultmidpunct}
{\mcitedefaultendpunct}{\mcitedefaultseppunct}\relax
\EndOfBibitem
\bibitem[S{\'a}ez-Bl{\'a}zquez \latin{et~al.}(2018)S{\'a}ez-Bl{\'a}zquez,
  Feist, Garc{\'\i}a-Vidal, and Fern{\'a}ndez-Dom{\'\i}nguez]{saez2018photon}
S{\'a}ez-Bl{\'a}zquez,~R.; Feist,~J.; Garc{\'\i}a-Vidal,~F.;
  Fern{\'a}ndez-Dom{\'\i}nguez,~A. Photon statistics in collective strong
  coupling: Nanocavities and microcavities. \emph{Phys. Rev. A} \textbf{2018},
  \emph{98}, 013839\relax
\mciteBstWouldAddEndPuncttrue
\mciteSetBstMidEndSepPunct{\mcitedefaultmidpunct}
{\mcitedefaultendpunct}{\mcitedefaultseppunct}\relax
\EndOfBibitem
\bibitem[Imamo\ifmmode~\bar{g}\else \={g}\fi{}lu
  \latin{et~al.}(1997)Imamo\ifmmode~\bar{g}\else \={g}\fi{}lu, Schmidt, Woods,
  and Deutsch]{PhysRevLett.79.1467}
Imamo\ifmmode~\bar{g}\else \={g}\fi{}lu,~A.; Schmidt,~H.; Woods,~G.;
  Deutsch,~M. Strongly Interacting Photons in a Nonlinear Cavity. \emph{Phys.
  Rev. Lett.} \textbf{1997}, \emph{79}, 1467--1470\relax
\mciteBstWouldAddEndPuncttrue
\mciteSetBstMidEndSepPunct{\mcitedefaultmidpunct}
{\mcitedefaultendpunct}{\mcitedefaultseppunct}\relax
\EndOfBibitem
\bibitem[Lang \latin{et~al.}(2011)Lang, Bozyigit, Eichler, Steffen, Fink,
  Abdumalikov, Baur, Filipp, da~Silva, Blais, and
  Wallraff]{PhysRevLett.106.243601}
Lang,~C.; Bozyigit,~D.; Eichler,~C.; Steffen,~L.; Fink,~J.~M.;
  Abdumalikov,~A.~A.; Baur,~M.; Filipp,~S.; da~Silva,~M.~P.; Blais,~A.;
  Wallraff,~A. Observation of Resonant Photon Blockade at Microwave Frequencies
  Using Correlation Function Measurements. \emph{Phys. Rev. Lett.}
  \textbf{2011}, \emph{106}, 243601\relax
\mciteBstWouldAddEndPuncttrue
\mciteSetBstMidEndSepPunct{\mcitedefaultmidpunct}
{\mcitedefaultendpunct}{\mcitedefaultseppunct}\relax
\EndOfBibitem
\bibitem[Rabl(2011)]{PhysRevLett.107.063601}
Rabl,~P. Photon Blockade Effect in Optomechanical Systems. \emph{Phys. Rev.
  Lett.} \textbf{2011}, \emph{107}, 063601\relax
\mciteBstWouldAddEndPuncttrue
\mciteSetBstMidEndSepPunct{\mcitedefaultmidpunct}
{\mcitedefaultendpunct}{\mcitedefaultseppunct}\relax
\EndOfBibitem
\bibitem[Snijders \latin{et~al.}(2018)Snijders, Frey, Norman, Flayac, Savona,
  Gossard, Bowers, van Exter, Bouwmeester, and
  L\"offler]{PhysRevLett.121.043601}
Snijders,~H.~J.; Frey,~J.~A.; Norman,~J.; Flayac,~H.; Savona,~V.;
  Gossard,~A.~C.; Bowers,~J.~E.; van Exter,~M.~P.; Bouwmeester,~D.;
  L\"offler,~W. Observation of the Unconventional Photon Blockade. \emph{Phys.
  Rev. Lett.} \textbf{2018}, \emph{121}, 043601\relax
\mciteBstWouldAddEndPuncttrue
\mciteSetBstMidEndSepPunct{\mcitedefaultmidpunct}
{\mcitedefaultendpunct}{\mcitedefaultseppunct}\relax
\EndOfBibitem
\bibitem[Vaneph \latin{et~al.}(2018)Vaneph, Morvan, Aiello, F\'echant, Aprili,
  Gabelli, and Est\`eve]{PhysRevLett.121.043602}
Vaneph,~C.; Morvan,~A.; Aiello,~G.; F\'echant,~M.; Aprili,~M.; Gabelli,~J.;
  Est\`eve,~J. Observation of the Unconventional Photon Blockade in the
  Microwave Domain. \emph{Phys. Rev. Lett.} \textbf{2018}, \emph{121},
  043602\relax
\mciteBstWouldAddEndPuncttrue
\mciteSetBstMidEndSepPunct{\mcitedefaultmidpunct}
{\mcitedefaultendpunct}{\mcitedefaultseppunct}\relax
\EndOfBibitem
\bibitem[Dhar \latin{et~al.}(2018)Dhar, Zens, Krimer, and
  Rotter]{PhysRevLett.121.133601}
Dhar,~H.~S.; Zens,~M.; Krimer,~D.~O.; Rotter,~S. Variational Renormalization
  Group for Dissipative Spin-Cavity Systems: Periodic Pulses of Nonclassical
  Photons from Mesoscopic Spin Ensembles. \emph{Phys. Rev. Lett.}
  \textbf{2018}, \emph{121}, 133601\relax
\mciteBstWouldAddEndPuncttrue
\mciteSetBstMidEndSepPunct{\mcitedefaultmidpunct}
{\mcitedefaultendpunct}{\mcitedefaultseppunct}\relax
\EndOfBibitem
\bibitem[Huang \latin{et~al.}(2018)Huang, Miranowicz, Liao, Nori, and
  Jing]{PhysRevLett.121.153601}
Huang,~R.; Miranowicz,~A.; Liao,~J.-Q.; Nori,~F.; Jing,~H. Nonreciprocal Photon
  Blockade. \emph{Phys. Rev. Lett.} \textbf{2018}, \emph{121}, 153601\relax
\mciteBstWouldAddEndPuncttrue
\mciteSetBstMidEndSepPunct{\mcitedefaultmidpunct}
{\mcitedefaultendpunct}{\mcitedefaultseppunct}\relax
\EndOfBibitem
\bibitem[Li \latin{et~al.}(2019)Li, Huang, Xu, Miranowicz, and Jing]{Jing2019}
Li,~B.; Huang,~R.; Xu,~X.; Miranowicz,~A.; Jing,~H. Nonreciprocal
  unconventional photon blockade in a spinning optomechanical system.
  \emph{Photon. Res.} \textbf{2019}, \emph{7}, 630--641\relax
\mciteBstWouldAddEndPuncttrue
\mciteSetBstMidEndSepPunct{\mcitedefaultmidpunct}
{\mcitedefaultendpunct}{\mcitedefaultseppunct}\relax
\EndOfBibitem
\bibitem[Liew and Savona(2010)Liew, and Savona]{PhysRevLett.104.183601}
Liew,~T. C.~H.; Savona,~V. Single Photons from Coupled Quantum Modes.
  \emph{Phys. Rev. Lett.} \textbf{2010}, \emph{104}, 183601\relax
\mciteBstWouldAddEndPuncttrue
\mciteSetBstMidEndSepPunct{\mcitedefaultmidpunct}
{\mcitedefaultendpunct}{\mcitedefaultseppunct}\relax
\EndOfBibitem
\bibitem[Radulaski \latin{et~al.}(2017)Radulaski, Fischer, Lagoudakis, Zhang,
  and Vu\ifmmode \check{c}\else \v{c}\fi{}kovi\ifmmode~\acute{c}\else
  \'{c}\fi{}]{PhysRevA.96.011801}
Radulaski,~M.; Fischer,~K.~A.; Lagoudakis,~K.~G.; Zhang,~J.~L.; Vu\ifmmode
  \check{c}\else \v{c}\fi{}kovi\ifmmode~\acute{c}\else \'{c}\fi{},~J. Photon
  blockade in two-emitter-cavity systems. \emph{Phys. Rev. A} \textbf{2017},
  \emph{96}, 011801\relax
\mciteBstWouldAddEndPuncttrue
\mciteSetBstMidEndSepPunct{\mcitedefaultmidpunct}
{\mcitedefaultendpunct}{\mcitedefaultseppunct}\relax
\EndOfBibitem
\bibitem[Bamba \latin{et~al.}(2011)Bamba, Imamo\ifmmode~\breve{g}\else
  \u{g}\fi{}lu, Carusotto, and Ciuti]{PhysRevA.83.021802}
Bamba,~M.; Imamo\ifmmode~\breve{g}\else \u{g}\fi{}lu,~A.; Carusotto,~I.;
  Ciuti,~C. Origin of strong photon antibunching in weakly nonlinear photonic
  molecules. \emph{Phys. Rev. A} \textbf{2011}, \emph{83}, 021802\relax
\mciteBstWouldAddEndPuncttrue
\mciteSetBstMidEndSepPunct{\mcitedefaultmidpunct}
{\mcitedefaultendpunct}{\mcitedefaultseppunct}\relax
\EndOfBibitem
\bibitem[Carre\~{n}o \latin{et~al.}()Carre\~{n}o, Casalengua, del Valle, and
  Laussy]{Carreno2016}
Carre\~{n}o,~J. C.~L.; Casalengua,~E.~Z.; del Valle,~E.; Laussy,~F.~P.
  Criterion for Single Photon Sources. \emph{arXiv:1610.06126v1} \relax
\mciteBstWouldAddEndPunctfalse
\mciteSetBstMidEndSepPunct{\mcitedefaultmidpunct}
{}{\mcitedefaultseppunct}\relax
\EndOfBibitem
\bibitem[Xu and Li(2013)Xu, and Li]{0953-4075-46-3-035502}
Xu,~X.-W.; Li,~Y.-J. Antibunching photons in a cavity coupled to an
  optomechanical system. \emph{J. Phys. B} \textbf{2013}, \emph{46},
  035502\relax
\mciteBstWouldAddEndPuncttrue
\mciteSetBstMidEndSepPunct{\mcitedefaultmidpunct}
{\mcitedefaultendpunct}{\mcitedefaultseppunct}\relax
\EndOfBibitem
\bibitem[Majumdar \latin{et~al.}(2012)Majumdar, Bajcsy, Rundquist, and
  Vu\ifmmode \check{c}\else \v{c}\fi{}kovi\ifmmode~\acute{c}\else
  \'{c}\fi{}]{PhysRevLett.108.183601}
Majumdar,~A.; Bajcsy,~M.; Rundquist,~A.; Vu\ifmmode \check{c}\else
  \v{c}\fi{}kovi\ifmmode~\acute{c}\else \'{c}\fi{},~J. Loss-Enabled
  Sub-Poissonian Light Generation in a Bimodal Nanocavity. \emph{Phys. Rev.
  Lett.} \textbf{2012}, \emph{108}, 183601\relax
\mciteBstWouldAddEndPuncttrue
\mciteSetBstMidEndSepPunct{\mcitedefaultmidpunct}
{\mcitedefaultendpunct}{\mcitedefaultseppunct}\relax
\EndOfBibitem
\bibitem[Zhou \latin{et~al.}(2015)Zhou, Shen, and Yi]{PhysRevA.92.023838}
Zhou,~Y.~H.; Shen,~H.~Z.; Yi,~X.~X. Unconventional photon blockade with
  second-order nonlinearity. \emph{Phys. Rev. A} \textbf{2015}, \emph{92},
  023838\relax
\mciteBstWouldAddEndPuncttrue
\mciteSetBstMidEndSepPunct{\mcitedefaultmidpunct}
{\mcitedefaultendpunct}{\mcitedefaultseppunct}\relax
\EndOfBibitem
\bibitem[Xu and Li(2014)Xu, and Li]{PhysRevA.90.033809}
Xu,~X.-W.; Li,~Y. Strong photon antibunching of symmetric and antisymmetric
  modes in weakly nonlinear photonic molecules. \emph{Phys. Rev. A}
  \textbf{2014}, \emph{90}, 033809\relax
\mciteBstWouldAddEndPuncttrue
\mciteSetBstMidEndSepPunct{\mcitedefaultmidpunct}
{\mcitedefaultendpunct}{\mcitedefaultseppunct}\relax
\EndOfBibitem
\bibitem[Santhosh \latin{et~al.}(2016)Santhosh, Bitton, Chuntonov, and
  Haran]{Santhosh2016}
Santhosh,~K.; Bitton,~O.; Chuntonov,~L.; Haran,~G. Vacuum Rabi splitting in a
  plasmonic cavity at the single quantum emitter limit. \emph{Nature
  Communications} \textbf{2016}, \emph{7}, 11823\relax
\mciteBstWouldAddEndPuncttrue
\mciteSetBstMidEndSepPunct{\mcitedefaultmidpunct}
{\mcitedefaultendpunct}{\mcitedefaultseppunct}\relax
\EndOfBibitem
\bibitem[Chikkaraddy \latin{et~al.}(2016)Chikkaraddy, de~Nijs, Benz, Barrow,
  Scherman, Rosta, Demetriadou, Fox, Hess, and Baumberg]{Chikkaraddy2016}
Chikkaraddy,~R.; de~Nijs,~B.; Benz,~F.; Barrow,~S.~J.; Scherman,~O.~A.;
  Rosta,~E.; Demetriadou,~A.; Fox,~P.; Hess,~O.; Baumberg,~J.~J.
  Single-molecule strong coupling at room temperature in plasmonic
  nanocavities. \emph{Nature} \textbf{2016}, \emph{535}, 127\relax
\mciteBstWouldAddEndPuncttrue
\mciteSetBstMidEndSepPunct{\mcitedefaultmidpunct}
{\mcitedefaultendpunct}{\mcitedefaultseppunct}\relax
\EndOfBibitem
\bibitem[Liu \latin{et~al.}(2017)Liu, Zhou, Yu, Zhang, Wang, Liu, Wei, Chen,
  and Wang]{PhysRevLett.118.237401}
Liu,~R.; Zhou,~Z.-K.; Yu,~Y.-C.; Zhang,~T.; Wang,~H.; Liu,~G.; Wei,~Y.;
  Chen,~H.; Wang,~X.-H. Strong Light-Matter Interactions in Single Open
  Plasmonic Nanocavities at the Quantum Optics Limit. \emph{Phys. Rev. Lett.}
  \textbf{2017}, \emph{118}, 237401\relax
\mciteBstWouldAddEndPuncttrue
\mciteSetBstMidEndSepPunct{\mcitedefaultmidpunct}
{\mcitedefaultendpunct}{\mcitedefaultseppunct}\relax
\EndOfBibitem
\bibitem[Ojambati \latin{et~al.}(2019)Ojambati, Chikkaraddy, Deacon, Horton,
  Kos, Turek, Keyser, and Baumberg]{ojambati2019quantum}
Ojambati,~O.~S.; Chikkaraddy,~R.; Deacon,~W.~D.; Horton,~M.; Kos,~D.;
  Turek,~V.~A.; Keyser,~U.~F.; Baumberg,~J.~J. Quantum electrodynamics at room
  temperature coupling a single vibrating molecule with a plasmonic nanocavity.
  \emph{Nat. Commun.} \textbf{2019}, \emph{10}, 1--7\relax
\mciteBstWouldAddEndPuncttrue
\mciteSetBstMidEndSepPunct{\mcitedefaultmidpunct}
{\mcitedefaultendpunct}{\mcitedefaultseppunct}\relax
\EndOfBibitem
\bibitem[Hood \latin{et~al.}(2000)Hood, Lynn, Doherty, Parkins, and
  Kimble]{Hood1447}
Hood,~C.~J.; Lynn,~T.~W.; Doherty,~A.~C.; Parkins,~A.~S.; Kimble,~H.~J. The
  Atom-Cavity Microscope: Single Atoms Bound in Orbit by Single Photons.
  \emph{Science} \textbf{2000}, \emph{287}, 1447--1453\relax
\mciteBstWouldAddEndPuncttrue
\mciteSetBstMidEndSepPunct{\mcitedefaultmidpunct}
{\mcitedefaultendpunct}{\mcitedefaultseppunct}\relax
\EndOfBibitem
\bibitem[Raimond \latin{et~al.}(2001)Raimond, Brune, and
  Haroche]{RevModPhys.73.565}
Raimond,~J.~M.; Brune,~M.; Haroche,~S. Manipulating quantum entanglement with
  atoms and photons in a cavity. \emph{Rev. Mod. Phys.} \textbf{2001},
  \emph{73}, 565--582\relax
\mciteBstWouldAddEndPuncttrue
\mciteSetBstMidEndSepPunct{\mcitedefaultmidpunct}
{\mcitedefaultendpunct}{\mcitedefaultseppunct}\relax
\EndOfBibitem
\bibitem[Chiorescu \latin{et~al.}(2004)Chiorescu, Bertet, Semba, Nakamura,
  Harmans, and Mooij]{Chiorescu2004}
Chiorescu,~I.; Bertet,~P.; Semba,~K.; Nakamura,~Y.; Harmans,~C. J. P.~M.;
  Mooij,~J.~E. Coherent dynamics of a flux qubit coupled to a harmonic
  oscillator. \emph{Nature} \textbf{2004}, \emph{431}, 159\relax
\mciteBstWouldAddEndPuncttrue
\mciteSetBstMidEndSepPunct{\mcitedefaultmidpunct}
{\mcitedefaultendpunct}{\mcitedefaultseppunct}\relax
\EndOfBibitem
\bibitem[Reithmaier \latin{et~al.}(2004)Reithmaier, S\c{e}k, L\"{o}ffler,
  Hofmann, Kuhn, Reitzenstein, Keldysh, Kulakovskii, Reinecke, and
  Forchel]{Reithmaier2004}
Reithmaier,~J.~P.; S\c{e}k,~G.; L\"{o}ffler,~A.; Hofmann,~C.; Kuhn,~S.;
  Reitzenstein,~S.; Keldysh,~L.~V.; Kulakovskii,~V.~D.; Reinecke,~T.~L.;
  Forchel,~A. Strong coupling in a single quantum dot-semiconductor microcavity
  system. \emph{Nature} \textbf{2004}, \emph{432}, 197\relax
\mciteBstWouldAddEndPuncttrue
\mciteSetBstMidEndSepPunct{\mcitedefaultmidpunct}
{\mcitedefaultendpunct}{\mcitedefaultseppunct}\relax
\EndOfBibitem
\bibitem[Baranov \latin{et~al.}(2018)Baranov, Wers\"{a}ll, Cuadra, Antosiewicz,
  and Shegai]{doi:10.1021/acsphotonics.7b00674}
Baranov,~D.~G.; Wers\"{a}ll,~M.; Cuadra,~J.; Antosiewicz,~T.~J.; Shegai,~T.
  Novel Nanostructures and Materials for Strong Light-Matter Interactions.
  \emph{ACS Photonics} \textbf{2018}, \emph{5}, 24--42\relax
\mciteBstWouldAddEndPuncttrue
\mciteSetBstMidEndSepPunct{\mcitedefaultmidpunct}
{\mcitedefaultendpunct}{\mcitedefaultseppunct}\relax
\EndOfBibitem
\bibitem[Soykal and Tahan(2013)Soykal, and Tahan]{PhysRevB.88.134511}
Soykal,~O.~O.; Tahan,~C. Toward engineered quantum many-body phonon systems.
  \emph{Phys. Rev. B} \textbf{2013}, \emph{88}, 134511\relax
\mciteBstWouldAddEndPuncttrue
\mciteSetBstMidEndSepPunct{\mcitedefaultmidpunct}
{\mcitedefaultendpunct}{\mcitedefaultseppunct}\relax
\EndOfBibitem
\bibitem[Xiong \latin{et~al.}(2020)Xiong, You, Bai, Png, Zhou, and
  Wu]{xiong2019ultrastrong}
Xiong,~X.; You,~J.-B.; Bai,~P.; Png,~C.~E.; Zhou,~Z.-K.; Wu,~L. Ultrastrong
  coupling in single plexcitonic nanocubes. \emph{Nanophotonics} \textbf{2020},
  \emph{9}, 257–66\relax
\mciteBstWouldAddEndPuncttrue
\mciteSetBstMidEndSepPunct{\mcitedefaultmidpunct}
{\mcitedefaultendpunct}{\mcitedefaultseppunct}\relax
\EndOfBibitem
\bibitem[Zengin \latin{et~al.}(2015)Zengin, Wers{\"a}ll, Nilsson, Antosiewicz,
  K{\"a}ll, and Shegai]{zengin2015realizing}
Zengin,~G.; Wers{\"a}ll,~M.; Nilsson,~S.; Antosiewicz,~T.~J.; K{\"a}ll,~M.;
  Shegai,~T. Realizing strong light-matter interactions between
  single-nanoparticle plasmons and molecular excitons at ambient conditions.
  \emph{Phys. Rev. Lett.} \textbf{2015}, \emph{114}, 157401\relax
\mciteBstWouldAddEndPuncttrue
\mciteSetBstMidEndSepPunct{\mcitedefaultmidpunct}
{\mcitedefaultendpunct}{\mcitedefaultseppunct}\relax
\EndOfBibitem
\bibitem[Cuadra \latin{et~al.}(2018)Cuadra, Baranov, Wers\"{a}ll, Verre,
  Antosiewicz, and Shegai]{cuadra2018observation}
Cuadra,~J.; Baranov,~D.~G.; Wers\"{a}ll,~M.; Verre,~R.; Antosiewicz,~T.~J.;
  Shegai,~T. Observation of tunable charged exciton polaritons in hybrid
  monolayer WS2- plasmonic nanoantenna system. \emph{Nano Lett.} \textbf{2018},
  \emph{18}, 1777--1785\relax
\mciteBstWouldAddEndPuncttrue
\mciteSetBstMidEndSepPunct{\mcitedefaultmidpunct}
{\mcitedefaultendpunct}{\mcitedefaultseppunct}\relax
\EndOfBibitem
\bibitem[Steck(2007)]{Ste07}
Steck,~D.~A. \emph{Quantum and Atom Optics}; 2007\relax
\mciteBstWouldAddEndPuncttrue
\mciteSetBstMidEndSepPunct{\mcitedefaultmidpunct}
{\mcitedefaultendpunct}{\mcitedefaultseppunct}\relax
\EndOfBibitem
\bibitem[Aghamalyan \latin{et~al.}(2019)Aghamalyan, You, Chu, Png, Krivitsky,
  and Kwek]{PhysRevA.100.053851}
Aghamalyan,~D.; You,~J.-B.; Chu,~H.-S.; Png,~C.~E.; Krivitsky,~L.; Kwek,~L.~C.
  Tunable quantum switch realized with a single
  $\mathrm{\ensuremath{\Lambda}}$-level atom coupled to the microtoroidal
  cavity. \emph{Phys. Rev. A} \textbf{2019}, \emph{100}, 053851\relax
\mciteBstWouldAddEndPuncttrue
\mciteSetBstMidEndSepPunct{\mcitedefaultmidpunct}
{\mcitedefaultendpunct}{\mcitedefaultseppunct}\relax
\EndOfBibitem
\bibitem[Berthel \latin{et~al.}(2015)Berthel, Mollet, Dantelle, Gacoin, Huant,
  and Drezet]{berthel2015photophysics}
Berthel,~M.; Mollet,~O.; Dantelle,~G.; Gacoin,~T.; Huant,~S.; Drezet,~A.
  Photophysics of single nitrogen-vacancy centers in diamond nanocrystals.
  \emph{Phys. Rev. B} \textbf{2015}, \emph{91}, 035308\relax
\mciteBstWouldAddEndPuncttrue
\mciteSetBstMidEndSepPunct{\mcitedefaultmidpunct}
{\mcitedefaultendpunct}{\mcitedefaultseppunct}\relax
\EndOfBibitem
\bibitem[Hamsen \latin{et~al.}(2017)Hamsen, Tolazzi, Wilk, and
  Rempe]{PhysRevLett.118.133604}
Hamsen,~C.; Tolazzi,~K.~N.; Wilk,~T.; Rempe,~G. Two-Photon Blockade in an
  Atom-Driven Cavity QED System. \emph{Phys. Rev. Lett.} \textbf{2017},
  \emph{118}, 133604\relax
\mciteBstWouldAddEndPuncttrue
\mciteSetBstMidEndSepPunct{\mcitedefaultmidpunct}
{\mcitedefaultendpunct}{\mcitedefaultseppunct}\relax
\EndOfBibitem
\bibitem[Klaas \latin{et~al.}(2018)Klaas, Schlottmann, Flayac, Laussy, Gericke,
  Schmidt, Helversen, Beyer, Brodbeck, Suchomel, H\"ofling, Reitzenstein, and
  Schneider]{PhysRevLett.121.047401}
Klaas,~M.; Schlottmann,~E.; Flayac,~H.; Laussy,~F.~P.; Gericke,~F.;
  Schmidt,~M.; Helversen,~M.~v.; Beyer,~J.; Brodbeck,~S.; Suchomel,~H.;
  H\"ofling,~S.; Reitzenstein,~S.; Schneider,~C. Photon-Number-Resolved
  Measurement of an Exciton-Polariton Condensate. \emph{Phys. Rev. Lett.}
  \textbf{2018}, \emph{121}, 047401\relax
\mciteBstWouldAddEndPuncttrue
\mciteSetBstMidEndSepPunct{\mcitedefaultmidpunct}
{\mcitedefaultendpunct}{\mcitedefaultseppunct}\relax
\EndOfBibitem
\bibitem[Deng \latin{et~al.}(2007)Deng, Solomon, Hey, Ploog, and
  Yamamoto]{PhysRevLett.99.126403}
Deng,~H.; Solomon,~G.~S.; Hey,~R.; Ploog,~K.~H.; Yamamoto,~Y. Spatial Coherence
  of a Polariton Condensate. \emph{Phys. Rev. Lett.} \textbf{2007}, \emph{99},
  126403\relax
\mciteBstWouldAddEndPuncttrue
\mciteSetBstMidEndSepPunct{\mcitedefaultmidpunct}
{\mcitedefaultendpunct}{\mcitedefaultseppunct}\relax
\EndOfBibitem
\bibitem[Kasprzak \latin{et~al.}(2008)Kasprzak, Richard, Baas, Deveaud,
  Andr\'e, Poizat, and Dang]{PhysRevLett.100.067402}
Kasprzak,~J.; Richard,~M.; Baas,~A.; Deveaud,~B.; Andr\'e,~R.; Poizat,~J.-P.;
  Dang,~L.~S. Second-Order Time Correlations within a Polariton Bose-Einstein
  Condensate in a CdTe Microcavity. \emph{Phys. Rev. Lett.} \textbf{2008},
  \emph{100}, 067402\relax
\mciteBstWouldAddEndPuncttrue
\mciteSetBstMidEndSepPunct{\mcitedefaultmidpunct}
{\mcitedefaultendpunct}{\mcitedefaultseppunct}\relax
\EndOfBibitem
\bibitem[Horikiri \latin{et~al.}(2010)Horikiri, Schwendimann, Quattropani,
  H\"ofling, Forchel, and Yamamoto]{PhysRevB.81.033307}
Horikiri,~T.; Schwendimann,~P.; Quattropani,~A.; H\"ofling,~S.; Forchel,~A.;
  Yamamoto,~Y. Higher order coherence of exciton-polariton condensates.
  \emph{Phys. Rev. B} \textbf{2010}, \emph{81}, 033307\relax
\mciteBstWouldAddEndPuncttrue
\mciteSetBstMidEndSepPunct{\mcitedefaultmidpunct}
{\mcitedefaultendpunct}{\mcitedefaultseppunct}\relax
\EndOfBibitem
\bibitem[A{\ss}mann \latin{et~al.}(2009)A{\ss}mann, Veit, Bayer, van~der Poel,
  and Hvam]{Assmann297}
A{\ss}mann,~M.; Veit,~F.; Bayer,~M.; van~der Poel,~M.; Hvam,~J.~M. Higher-Order
  Photon Bunching in a Semiconductor Microcavity. \emph{Science} \textbf{2009},
  \emph{325}, 297--300\relax
\mciteBstWouldAddEndPuncttrue
\mciteSetBstMidEndSepPunct{\mcitedefaultmidpunct}
{\mcitedefaultendpunct}{\mcitedefaultseppunct}\relax
\EndOfBibitem
\bibitem[M{\o}lmer \latin{et~al.}(1993)M{\o}lmer, Castin, and
  Dalibard]{Molmer:93}
M{\o}lmer,~K.; Castin,~Y.; Dalibard,~J. Monte Carlo wave-function method in
  quantum optics. \emph{J. Opt. Soc. Am. B} \textbf{1993}, \emph{10},
  524--538\relax
\mciteBstWouldAddEndPuncttrue
\mciteSetBstMidEndSepPunct{\mcitedefaultmidpunct}
{\mcitedefaultendpunct}{\mcitedefaultseppunct}\relax
\EndOfBibitem
\bibitem[Xu \latin{et~al.}(2018)Xu, Xiong, Wu, Ren, Png, Guo, Gong, and
  Xiao]{xu2018quantum}
Xu,~D.; Xiong,~X.; Wu,~L.; Ren,~X.-F.; Png,~C.~E.; Guo,~G.-C.; Gong,~Q.;
  Xiao,~Y.-F. Quantum plasmonics: new opportunity in fundamental and applied
  photonics. \emph{Adv. Opt. Photonics} \textbf{2018}, \emph{10},
  703--756\relax
\mciteBstWouldAddEndPuncttrue
\mciteSetBstMidEndSepPunct{\mcitedefaultmidpunct}
{\mcitedefaultendpunct}{\mcitedefaultseppunct}\relax
\EndOfBibitem
\bibitem[Zhou \latin{et~al.}(2019)Zhou, Liu, Bao, Wu, Png, Wang, and
  Qiu]{zhou2019quantum}
Zhou,~Z.-K.; Liu,~J.; Bao,~Y.; Wu,~L.; Png,~C.~E.; Wang,~X.-H.; Qiu,~C.-W.
  Quantum plasmonics get applied. \emph{Prog. Quantum Electron.} \textbf{2019},
  \emph{65}, 1--20\relax
\mciteBstWouldAddEndPuncttrue
\mciteSetBstMidEndSepPunct{\mcitedefaultmidpunct}
{\mcitedefaultendpunct}{\mcitedefaultseppunct}\relax
\EndOfBibitem
\bibitem[Qian \latin{et~al.}(2018)Qian, Wu, Song, Peng, Xie, Yang, Xiao, Steer,
  Thayne, Tang, \latin{et~al.} others]{qian2018two}
Qian,~C.; Wu,~S.; Song,~F.; Peng,~K.; Xie,~X.; Yang,~J.; Xiao,~S.;
  Steer,~M.~J.; Thayne,~I.~G.; Tang,~C., \latin{et~al.}  Two-photon Rabi
  splitting in a coupled system of a nanocavity and exciton complexes.
  \emph{Phys. Rev. Lett.} \textbf{2018}, \emph{120}, 213901\relax
\mciteBstWouldAddEndPuncttrue
\mciteSetBstMidEndSepPunct{\mcitedefaultmidpunct}
{\mcitedefaultendpunct}{\mcitedefaultseppunct}\relax
\EndOfBibitem
\bibitem[Schwartz \latin{et~al.}(2011)Schwartz, Hutchison, Genet, and
  Ebbesen]{schwartz2011reversible}
Schwartz,~T.; Hutchison,~J.~A.; Genet,~C.; Ebbesen,~T.~W. Reversible switching
  of ultrastrong light-molecule coupling. \emph{Phys. Rev. Lett.}
  \textbf{2011}, \emph{106}, 196405\relax
\mciteBstWouldAddEndPuncttrue
\mciteSetBstMidEndSepPunct{\mcitedefaultmidpunct}
{\mcitedefaultendpunct}{\mcitedefaultseppunct}\relax
\EndOfBibitem
\bibitem[Greentree \latin{et~al.}(2006)Greentree, Tahan, Cole, and
  Hollenberg]{Greentree2006}
Greentree,~A.~D.; Tahan,~C.; Cole,~J.~H.; Hollenberg,~L. C.~L. Quantum phase
  transitions of light. \emph{Nature} \textbf{2006}, \emph{2}, 856\relax
\mciteBstWouldAddEndPuncttrue
\mciteSetBstMidEndSepPunct{\mcitedefaultmidpunct}
{\mcitedefaultendpunct}{\mcitedefaultseppunct}\relax
\EndOfBibitem
\bibitem[Hartmann \latin{et~al.}(2006)Hartmann, Brand\~{a}o, and
  Plenio]{Hartmann2006}
Hartmann,~M.~J.; Brand\~{a}o,~F. G. S.~L.; Plenio,~M.~B. Strongly interacting
  polaritons in coupled arrays of cavities. \emph{Nature Physics}
  \textbf{2006}, \emph{2}, 849\relax
\mciteBstWouldAddEndPuncttrue
\mciteSetBstMidEndSepPunct{\mcitedefaultmidpunct}
{\mcitedefaultendpunct}{\mcitedefaultseppunct}\relax
\EndOfBibitem
\bibitem[You \latin{et~al.}(2014)You, Yang, Xu, Chan, and
  Oh]{PhysRevB.90.195112}
You,~J.-B.; Yang,~W.~L.; Xu,~Z.-Y.; Chan,~A.~H.; Oh,~C.~H. Phase transition of
  light in circuit-QED lattices coupled to nitrogen-vacancy centers in diamond.
  \emph{Phys. Rev. B} \textbf{2014}, \emph{90}, 195112\relax
\mciteBstWouldAddEndPuncttrue
\mciteSetBstMidEndSepPunct{\mcitedefaultmidpunct}
{\mcitedefaultendpunct}{\mcitedefaultseppunct}\relax
\EndOfBibitem
\bibitem[Xiao \latin{et~al.}(2012)Xiao, Liu, Li, Chen, Li, and
  Gong]{xiao2012strongly}
Xiao,~Y.-F.; Liu,~Y.-C.; Li,~B.-B.; Chen,~Y.-L.; Li,~Y.; Gong,~Q. Strongly
  enhanced light-matter interaction in a hybrid photonic-plasmonic resonator.
  \emph{Phys. Rev. A} \textbf{2012}, \emph{85}, 031805\relax
\mciteBstWouldAddEndPuncttrue
\mciteSetBstMidEndSepPunct{\mcitedefaultmidpunct}
{\mcitedefaultendpunct}{\mcitedefaultseppunct}\relax
\EndOfBibitem
\bibitem[Peng \latin{et~al.}(2017)Peng, Liu, Xu, Cao, Lu, Gong, Xiao,
  \latin{et~al.} others]{peng2017enhancing}
Peng,~P.; Liu,~Y.-C.; Xu,~D.; Cao,~Q.-T.; Lu,~G.; Gong,~Q.; Xiao,~Y.-F.,
  \latin{et~al.}  Enhancing coherent light-matter interactions through
  microcavity-engineered plasmonic resonances. \emph{Phys. Rev. Lett.}
  \textbf{2017}, \emph{119}, 233901\relax
\mciteBstWouldAddEndPuncttrue
\mciteSetBstMidEndSepPunct{\mcitedefaultmidpunct}
{\mcitedefaultendpunct}{\mcitedefaultseppunct}\relax
\EndOfBibitem
\bibitem[Deng \latin{et~al.}(2017)Deng, Li, and Qin]{Deng:17}
Deng,~W.-W.; Li,~G.-X.; Qin,~H. Photon blockade via quantum interference in a
  strong coupling qubit-cavity system. \emph{Opt. Express} \textbf{2017},
  \emph{25}, 6767--6783\relax
\mciteBstWouldAddEndPuncttrue
\mciteSetBstMidEndSepPunct{\mcitedefaultmidpunct}
{\mcitedefaultendpunct}{\mcitedefaultseppunct}\relax
\EndOfBibitem
\end{mcitethebibliography}







\end{document}